\documentclass[aip,jcp,reprint,twocolumn,superscriptaddress,showpacs,floatfix]{revtex4-1}

\usepackage{mathrsfs}
\usepackage{amssymb}
\usepackage{amsmath}
\usepackage{dcolumn}
\usepackage{bm}
\usepackage{gensymb}
\usepackage{physics}
\usepackage{float}
\usepackage{multirow}
\usepackage{color}
\usepackage{tabularx}
\usepackage{hyperref}
\newcommand{\Cltwo}{Cl$_{2}$}
\newcommand{\Brtwo}{Br$_{2}$}
\newcommand{\Cltwoplus}{$\mathrm{(Cl_{2})}^{2+}$}
\newcommand{\Brtwoplus}{$\mathrm{(Br_{2})}^{2+}$}
\newcommand{\HBrtwoplus}{$\mathrm{(HBr)}^{2+}$}
\newcommand{\water}{H$_{2}$O}
\newcommand{\watertwoplus}{$\mathrm{(H_{2}O)^{2+}}$}
\newcommand{\watersinglet}{$a\: ^{1}\mathrm{A}_{1}$}
\newcommand{\watertriplet}{$X\: ^{3}\mathrm{B}_{1}$}
\newcommand{\methane}{CH$_{4}$}
\newcommand{\methanesinglet}{$a\: ^{1}\mathrm{E}$}
\newcommand{\methanetriplet}{$X\: ^{3}\mathrm{T}_{1}$}
\newcommand{\methanetwoplus}{$\mathrm{(C H_{4})^{2+}}$}
\newcommand{\bntwoplus}{$\mathrm{(BN)^{2+}}$}
\newcommand{\bnsinglet}{$a\: ^{1}\Delta$}
\newcommand{\bntriplet}{$X\: ^{3}\Sigma^{-}$}
\definecolor{red}{rgb}{1,0,0}
\begin{document}

\title{Double Ionization Potential Equation-of-Motion Coupled-Cluster Approach
with Full Inclusion of 4-Hole--2-Particle Excitations and Three-Body Clusters}

\author{Karthik Gururangan}
\email[e-mail: ]{gururang@msu.edu}
\affiliation{Department of Chemistry,
Michigan State University, East Lansing, MI 48824, USA}
\author{Achintya Kumar Dutta}
\email[e-mail: ]{achintya@chem.iitb.ac.in}
\affiliation{Department of Chemistry, 
Indian Institute of Technology Bombay, Powai, Mumbai 400076, India}
\author{Piotr Piecuch}
\thanks{Corresponding author}
\email[e-mail: ]{piecuch@chemistry.msu.edu}
\affiliation{Department of Chemistry,
Michigan State University, East Lansing, MI 48824, USA}
\affiliation{Department of Physics and Astronomy,
Michigan State University, East Lansing, MI 48824, USA}

\begin{abstract}
The double ionization potential (DIP) equation-of-motion (EOM) coupled-cluster (CC) method
with a full treatment of 4-hole--2-particle (4$h$-2$p$) correlations and triply excited
clusters, abbreviated as DIP-EOMCCSDT(4$h$-2$p$), and its approximate form called
DIP-EOMCCSD(T)(a)(4$h$-2$p$) have been formulated and implemented in the open-source CCpy package
available on GitHub.
The resulting codes work with both nonrelativistic and spin-free scalar-relativistic Hamiltonians.
By examining the DIPs of
a few small molecules,
for which accurate
reference
data are
available, we demonstrate that the DIP-EOMCCSDT(4$h$-2$p$) and DIP-EOMCCSD(T)(a)(4$h$-2$p$)
approaches improve the results obtained using the DIP-EOMCC methods truncated at 3$h$-1$p$
or 4$h$-2$p$ excitations on top of the CC calculations with singles and doubles.
\end{abstract}

\maketitle
\twocolumngrid
The single-reference coupled-cluster (CC) theory\cite{Coester:1958,Coester:1960,cizek1,cizek2,cizek4}
and its equation-of-motion (EOM) extensions to electronically
excited\cite{emrich,emrich2,eomcc1,eomcc2,eomcc3} and electron attached and ionized states
\cite{eaccsd1,eaccsd2,hirataeaip,eaccsdt1,gour1,gour2,gour3,%
ipccsd2,ipccsd3,ipccsd4,ipccsd1,ipccsdt1,ipccsdt2,ipccsdt-3,hirata_ea_ip,%
dipea1,steom2,dipea2,dipea3,dip-stanton,dipea5,dipea6,kus-krylov-2011,kus-krylov-2012,%
jspp-dea-dip-2013,jspp-dea-dip-2014,jspp-dea-dip-2017,jspp-dea-dip-2021,dipea7,tea-eomcc}
have become preeminent methods of quantum chemistry. In this Communication,
we focus on the double ionization potential (DIP) EOMCC framework, which allows one
to directly determine the ground and excited states of doubly ionized molecular species
and which is useful in many applications, such as Auger electron spectroscopy,
\cite{ghosh-auger,krylov-auger-1,krylov-auger-2,auger-benzene} singlet--triplet gaps in biradicals,
\cite{kus-krylov-2011,kus-krylov-2012,jspp-dea-dip-2013,jspp-dea-dip-2014}
and strong-field-induced chemical reactivity.\cite{nat-commun-h3plus}

In the DIP-EOMCC formalism, the ground ($\mu = 0$) and excited ($\mu > 0$) states of the 
target ($N-2$)-electron system are expressed as
\begin{equation}
|\Psi_{\mu}^{(N-2)}\rangle = R_{\mu}^{(-2)} |\Psi^{(N)}\rangle,    
\label{dipwf}
\end{equation}
where the doubly ionizing operator
\begin{equation}
R_{\mu}^{(-2)} = \sum_{n=0}^{M_{R}} R_{\mu,(n+2)h\mbox{-}np},
\label{Rop}
\end{equation}
with $R_{\mu,(n+2)h\mbox{-}np}$ representing its $(n+2)$-hole--$n$-particle [$(n+2)h$--$np$]
components, removes two electrons from the CC ground state
\begin{equation}
|\Psi^{(N)}\rangle = e^{T} |\Phi\rangle
\label{ccwf}
\end{equation}
of the underlying $N$-electron species, in which
\begin{equation}
T = \sum_{n=1}^{M_{T}} T_{n}
\label{Top}
\end{equation}
is the cluster operator and $|\Phi\rangle$ is the reference determinant
that serves as a Fermi vacuum.
The many-body components of the $T$ and $R_{\mu}^{(-2)}$ operators are given by
\begin{equation}
T_{n} = \sum_{\substack{i_{1}<\cdots<i_{n} \\ a_{1}<\cdots<a_{n}}}
t_{a_{1}\ldots a_{n}}^{i_{1}\ldots i_{n}} a^{a_{1}}\ldots a^{a_{n}} a_{i_{n}}\ldots a_{i_{1}}
\label{Tn}
\end{equation}
and
\begin{align}
R_{\mu,(n+2)h\mbox{-}np} = &\sum_{\substack{k_{1}<\cdots<k_{n}<i<j \\ c_{1} < \cdots < c_{n}}} r_{\phantom{ab}c_{1}\ldots c_{n}}^{ijk_{1}\ldots k_{n}}(\mu)  \nonumber \\
&\times a^{c_{1}}\ldots a^{c_{n}}a_{k_{n}}\ldots a_{k_{1}}a_{j}a_{i},
\label{Rn}
\end{align}
respectively, where, as usual, indices $i,j, \, \ldots$ ($a,b, \, \ldots$) denote the 
spinorbitals that are occupied (unoccupied) in $|\Phi\rangle$ and $a^{p}$ ($a_{p}$) 
represents the fermionic creation (annihilation) operator 
associated with the spinorbital $|p\rangle$. $M_{R}$ and $M_{T}$ control the 
DIP-EOMCC theory level, with  $M_{R} = N -2$ defining exact calculations and $M_{R} < N - 2$ and
$M_{T} \leq N$ leading to DIP-EOMCC approximations.

The DIP-EOMCC approaches that have been implemented so far include the basic
DIP-EOMCCSD(3$h$-1$p$) method,
\cite{dipea2,dipea3,dipea5,kus-krylov-2011,kus-krylov-2012,%
jspp-dea-dip-2013,jspp-dea-dip-2014}
defined by $M_{R} = 1$ and $M_{T} = 2$,
which describes 2$h$ and 3$h$-1$p$ correlations on top of the CC calculations
with singles and doubles (CCSD),\cite{ccsd,ccsd2,ccsdfritz,osaccsd} its
higher-level DIP-EOMCCSD(4$h$-2$p$) extension\cite{jspp-dea-dip-2013,jspp-dea-dip-2014}
corresponding to $M_{R} = 2$ and $M_{T} = 2$,
which also describes 4$h$-2$p$ correlations on top of CCSD, and the DI-EOMCCSDT scheme\cite{dipea5}
corresponding to $M_{R} = 1$ and $M_{T} = 3$, which accounts for 2$h$ and 3$h$-1$p$ correlations
on top of the CC calculations with singles, doubles, and triples (CCSDT).\cite{ccfullt,ccfullt2}
While all of these methods and the various approximations to them aimed at reducing computational costs are
useful tools for determining DIP energies in molecular systems and singlet--triplet gaps in certain biradicals,
our recent numerical tests, including the DIPs of
a few small molecules
examined
in this Communication, indicate that the explicit incorporation of 4$h$-2$p$ correlations,
needed to achieve a highly accurate description,\cite{jspp-dea-dip-2013,jspp-dea-dip-2014}
may not be well balanced with the CCSD treatment of the underlying $N$-electron
species, resulting in some cases in loss of accuracy compared to the basic DIP-EOMCCSD(3$h$-1$p$) level.

The main goal of this work is to enrich the existing arsenal of DIP-EOMCC methods and, in particular, address the
above concerns by examining the high-level DIP-EOMCC approach with a full treatment of both 4$h$-2$p$ and
$T_{3}$ correlations, corresponding to $M_{R} = 2$ and $M_{T} = 3$ and abbreviated as DIP-EOMCCSDT(4$h$-2$p$).
The present study describes the efficient formulation and computer implementation of full DIP-EOMCCSDT(4$h$-2$p$)
and its less expensive DIP-EOMCCSD(T)(a)(4$h$-2$p$) approximation, including the factorized and programmable
expressions that define them. We illustrate the performance of both methods, as coded for nonrelativistic and
spin-free scalar-relativistic Hamiltonians in the open-source CCpy package\cite{CCpy-GitHub} available on GitHub,
by calculating the vertical DIPs of
\water{}, \methane{}, BN,
\Cltwo{}, \Brtwo{}, and HBr and comparing the results with those obtained
using the DIP-EOMCCSD(3$h$-1$p$) and DIP-EOMCCSD(4$h$-2$p$) approaches and the
reliable, high-accuracy, reference
data.

The key step of any DIP-EOMCC calculation is a diagonalization of the similarity-transformed Hamiltonian
$\overline{H}_{N} = e^{-T} H_{N} e^{T} = (H_{N} e^{T})_C$ associated with the $N$-electron CC ground state
in the relevant ($N - 2$)-electron subspace of the Fock space corresponding to the content of the ionizing
operator $R_{\mu}^{(-2)}$. Here, subscript $C$ designates the connected operator product and
$H_{N} = H - \langle \Phi | H | \Phi \rangle = F_{N} + V_{N}$ is the electronic Hamiltonian in the normal-ordered
form relative to $|\Phi\rangle$, with $F_{N}$ and $V_{N}$ representing its Fock and two-electron interaction
components. In the case of the cluster and ionizing operators defined by Eqs.\ (\ref{Top}) and (\ref{Rop}),
respectively, the basis states that span the ($N-2$)-electron subspace of the Fock space used in the DIP-EOMCC
calculations are $|\Phi_{ij}\rangle = a_{j}a_{i}|\Phi\rangle$ and
$|\Phi_{ijk_{1}\ldots k_{n}}^{\phantom{ab}c_{1}\ldots c_{n}}\rangle =
a^{c_{1}}\ldots a^{c_{n}}a_{k_{n}}\ldots a_{k_{1}}a_{j}a_{i}|\Phi\rangle$, $n = 1,\ldots,M_{R}$. 
Assuming that $M_{R} \leq M_{T}$ (a condition required for retaining size intensivity of the resulting
double ionization energies\cite{jspp-dea-dip-2013,jspp-dea-dip-2014,jspp-dea-dip-2021}), the DIP-EOMCC
eigenvalue problem is given by
\begin{equation}
(\overline{H}_{N,\text{open}} R_{\mu}^{(-2)})_C |\Phi\rangle = \omega_{\mu}^{(N - 2)} R_{\mu}^{(-2)}|\Phi\rangle,
\label{dipeqn}
\end{equation}
where $\overline{H}_{N,\text{open}}$ refers to the diagrams of $\overline{H}_{N}$ containing 
external fermion lines and $\omega_{\mu}^{(N - 2)} = E_{\mu}^{(N - 2)} - E_{0}^{(N)}$ is the 
vertical DIP energy representing the difference between the total energy of the ground ($\mu = 0$) or
excited ($\mu > 0$) state of the ($N - 2$)-electron target system, denoted as $E_{\mu}^{(N - 2)}$, and
the ground-state CC energy of the underlying $N$-electron species,
$E_{0}^{(N)} = \langle \Phi | H | \Phi \rangle + \langle \Phi | \overline{H}_{N} | \Phi \rangle$. 
In practice, including this work, the solutions of Eq.\ (\ref{dipeqn}) are obtained using the
Hirao--Nakatsuji generalization\cite{hirao} of the Davidson diagonalization algorithm\cite{dav} to
non-Hermitian Hamiltonians.

In the DIP-EOMCCSDT(4$h$-2$p$) approach developed in this work, the CCSDT 
similarity-transformed Hamiltonian, designated as $\overline{H}_{N}^{(\text{CCSDT})}$,
is diagonalized in the $(N-2)$-electron subspace of the Fock space
spanned by $|\Phi_{ij}\rangle$, $|\Phi_{ijk}^{\phantom{ab}c}\rangle$, 
and $|\Phi_{ijkl}^{\phantom{ab}cd}\rangle$.  
The programmable factorized expressions for the required projections of the
left-hand-side of Eq.\ (\ref{dipeqn}), with $T$ truncated at $T_{3}$ and $R_{\mu}^{(-2)}$
truncated at $R_{\mu,4h\mbox{-}2p}$, as implemented in CCpy, are
\begin{widetext}
\begin{equation}
\langle \Phi_{ij} | (\overline{H}_{N,\mathrm{open}}^{(\text{CCSDT})} R_{\mu}^{(-2)})_C | \Phi \rangle 
= \mathscr{A}^{ij}[
-\bar{h}_{m}^{i} r_{}^{mj}(\mu)
+ \tfrac{1}{4} \bar{h}_{mn}^{ij}r_{}^{mn}(\mu) 
+ \tfrac{1}{2} \bar{h}_{m}^{e}r_{\phantom{ab}e}^{ijm}(\mu) 
- \tfrac{1}{2} \bar{h}_{mn}^{if}r_{\phantom{ab}f}^{mjn}(\mu)
+ \tfrac{1}{8} \bar{h}_{mn}^{ef}r_{\phantom{ab}ef}^{ijmn}(\mu)],
\label{eq2h}
\end{equation}
\begin{align}
\langle \Phi_{ijk}^{\phantom{ab}c} | (\overline{H}_{N,\mathrm{open}}^{(\text{CCSDT})} R_{\mu}^{(-2)})_C | \Phi \rangle 
=&\mathscr{A}^{ijk}[
\tfrac{1}{2} {I^{\prime}}^{ie}(\mu) t_{ec}^{jk} 
- \tfrac{1}{2} \bar{h}_{cm}^{ki} r_{}^{mj}(\mu)
+ \tfrac{1}{6} \bar{h}_{c}^{e} r_{\phantom{ab}e}^{ijk}(\mu)
- \tfrac{1}{2} \bar{h}_{m}^{k} r_{\phantom{ab}c}^{ijm}(\mu)
+ \tfrac{1}{4} \bar{h}_{mn}^{ij} r_{\phantom{ab}c}^{mnk}(\mu)
\nonumber \\
&+ \tfrac{1}{2} \bar{h}_{cm}^{ke} r_{\phantom{ab}e}^{ijm}(\mu)
+ \tfrac{1}{6} \bar{h}_{m}^{e} r_{\phantom{ab}ce}^{ijkm}(\mu)
- \tfrac{1}{4} \bar{h}_{mn}^{kf} r_{\phantom{ab}cf}^{ijmn}(\mu)
+ \tfrac{1}{12} \bar{h}_{cn}^{ef} r_{\phantom{ab}ef}^{ijkn}(\mu)
\nonumber \\
&+ \tfrac{1}{12} I^{ef} t_{efc}^{ijk}],
\label{eq3h1p}
\end{align}
and
\begin{align}
\langle \Phi_{ijkl}^{\phantom{ab}cd} | (\overline{H}_{N,\mathrm{open}}^{(\text{CCSDT})} R_{\mu}^{(-2)})_C | \Phi \rangle 
= &\mathscr{A}^{ijkl}\mathscr{A}_{cd}[
\tfrac{1}{12} \bar{h}_{dc}^{le} r_{\phantom{ab}e}^{ijk}(\mu)
- \tfrac{1}{4} \bar{h}_{dm}^{lk} r_{\phantom{ab}c}^{ijm}(\mu)
- \tfrac{1}{12} I_{\phantom{ab}m}^{ijk}(\mu) t_{cd}^{ml}
+ \tfrac{1}{4} I_{\phantom{ab}c}^{ije}(\mu) t_{ed}^{kl}
+ \tfrac{1}{24} \bar{h}_{d}^{e} r_{\phantom{ab}ce}^{ijkl}(\mu)
\nonumber \\
&- \tfrac{1}{12} \bar{h}_{m}^{i} r_{\phantom{ab}cd}^{mjkl}(\mu)
+ \tfrac{1}{16} \bar{h}_{mn}^{ij} r_{\phantom{ab}cd}^{mnkl}(\mu)
+ \tfrac{1}{96} \bar{h}_{cd}^{ef} r_{\phantom{ab}ef}^{ijkl}(\mu)
+ \tfrac{1}{6} \bar{h}_{dm}^{le} r_{\phantom{ab}ce}^{ijkm}(\mu)
\nonumber \\
&+ \tfrac{1}{12} I^{ie}(\mu) t_{ecd}^{jkl}
+ \tfrac{1}{8} I_{\phantom{ab}e}^{ijm}(\mu) t_{cde}^{klm}
+ \tfrac{1}{12} I_{\phantom{ab}c}^{efk}(\mu) t_{efd}^{ijl}],
\label{eq4h2p}
\end{align}
\end{widetext}
where we use the Einstein summation convention over repeated
upper and lower indices and $\mathscr{A}^{pq} = \mathscr{A}_{pq} = 1 - (pq)$,
$\mathscr{A}^{pqr} = \mathscr{A}^{p/qr}\mathscr{A}^{qr}$, 
and $\mathscr{A}^{pqrs} = \mathscr{A}^{p/qrs}\mathscr{A}^{qrs}$, 
with $\mathscr{A}^{p/qr} = 1 - (pq) - (pr)$ and
$\mathscr{A}^{p/qrs} = 1 - (pq) - (pr) - (ps)$,
are index antisymmetrizers. The expressions for
the one-body ($\bar{h}_{p}^{q}$) and two-body ($\bar{h}_{pq}^{rs}$) 
components of the similarity-transformed Hamiltonian
as well as the additional intermediates entering Eqs.\ (\ref{eq3h1p}) and (\ref{eq4h2p})
are provided in Tables \ref{table1} and \ref{table2}.
Equations (\ref{eq2h})--(\ref{eq4h2p}) imply that the diagonalization step of
DIP-EOMCCSDT(4$h$-2$p$) has computational costs identical to those characterizing
DIP-EOMCCSD(4$h$-2$p$), which scale as $n_{o}^4 n_{u}^4$
or $\mathscr{N}^8$, where $n_{o}$ ($n_{u}$) is the number of occupied (unoccupied)
orbitals in $|\Phi\rangle$ and $\mathscr{N}$ is a measure of the system size.
However, the overall computational effort associated with the DIP-EOMCCSDT(4$h$-2$p$)
approach is considerably higher than that of DIP-EOMCCSD(4$h$-2$p$) since
in the DIP-EOMCCSDT(4$h$-2$p$) case, one also has to solve the CCSDT equations 
for the underlying $N$-electron species, which involve $n_{o}^3 n_{u}^5$ steps,
as opposed to the much less expensive $n_{o}^2 n_{u}^4$ ($\mathscr{N}^6$) steps of
CCSD.

Given the high computational costs of the DIP-EOMCCSDT(4$h$-2$p$) method,
we also consider the more practical DIP-EOMCCSD(T)(a)(4$h$-2$p$) scheme,
in which we adopt the M{\o}ller-Plesset (MP) partitioning of the Hamiltonian
and, following Ref.\ \onlinecite{eomccsdta}, incorporate the leading $T_{3}$ correlation
effects by correcting the $T_{1}$ and $T_{2}$ clusters obtained with CCSD
using the formulas
\begin{equation}
\tilde{T}_{1} |\Phi\rangle = T_{1} |\Phi\rangle + D_{1}^{-1}(V_{N} \tilde{T}_{3})_C |\Phi\rangle
\label{t1pert}
\end{equation}
and 
\begin{equation}
\tilde{T}_{2} |\Phi\rangle = T_{2} |\Phi\rangle + D_{2}^{-1}(H_{N} \tilde{T}_{3})_C |\Phi\rangle,
\label{t2pert}
\end{equation}
where $D_{1}$ and $D_{2}$ are the usual MP denominators for singles and doubles and
\begin{equation}
\tilde{T}_{3} |\Phi\rangle = D_{3}^{-1}(V_{N} T_{2})_C |\Phi\rangle
\label{t3pert}
\end{equation}
is the lowest-order approximation to $T_{3}$, with $D_{3}$ representing the MP denominator for triples.
Once $\tilde{T}_{1}$, $\tilde{T}_{2}$, and $\tilde{T}_{3}$ are determined via
Eqs.\ (\ref{t1pert})--(\ref{t3pert}), the resulting CCSD(T)(a) similarity-transformed Hamiltonian,
constructed using the recipe described in Ref.\ \onlinecite{eomccsdta},
is diagonalized in the same way as $\overline{H}^{(\text{CCSDT})}_{N}$
in DIP-EOMCCSDT(4$h$-2$p$) with the help of Eqs.\ (\ref{eq2h})--(\ref{eq4h2p}).
By eliminating the need for performing CCSDT calculations, the most expensive steps
of DIP-EOMCCSD(T)(a)(4$h$-2$p$) scale as $n_{o}^4 n_{u}^4$ rather than $n_{o}^3 n_{u}^5$.

To illustrate the performance of the DIP-EOMCCSDT(4$h$-2$p$) and DIP-EOMCCSD(T)(a)(4$h$-2$p$) methods,
as implemented for nonrelativistic and spin-free scalar-relativistic Hamiltonians in the
open-source CCpy package,\cite{CCpy-GitHub} we applied them, along with their DIP-EOMCCSD(3$h$-1$p$) and
DIP-EOMCCSD(4$h$-2$p$) counterparts, available in CCpy as well, to the vertical DIPs of
two sets of small molecules, for which highly accurate reference data can be found in
the literature. The first set, which is the primary focus on our discussion below, consisted of the
\water{}, \methane{}, and BN molecules, as described by the aug-cc-pVTZ basis set,\cite{ccpvnz,augccpvnz}
where we compared our DIP-EOMCC DIPs corresponding to the lowest singlet and triplet states of the
\watertwoplus{}, \methanetwoplus{}, and \bntwoplus{} dications with their near-exact counterparts obtained
in Ref.\ \onlinecite{loos-ppbse} with the configuration interaction (CI) approach using perturbative selection
made iteratively, abbreviated as CIPSI,\cite{sci_3,cipsi_1,cipsi_2} extrapolated to the full CI limit.
We also calculated the vertical DIPs of
the \Cltwo{}, \Brtwo{}, and HBr molecules
corresponding to
the triplet ground states and low-lying singlet states of
\Cltwoplus{}, \Brtwoplus{}, and \HBrtwoplus{}, comparing our DIP-EOMCC results with the
reliable experimental data reported
in Refs.\ \onlinecite{dip_cl2,dip_br2,dip_hbr}.
In this case, to
obtain insights into the
convergence of our calculated DIP values with the basis set, we used
the cc-pVTZ and cc-pVQZ bases.\cite{ccpvnz,ccpvnz3,ccpvnz9}
Following Ref.\ \onlinecite{loos-ppbse}, the
equilibrium geometries of the ground-state \water{}, \methane{}, and BN molecules used in our DIP-EOMCC computations were
taken from Ref.\ \onlinecite{loos-ip}, where they were optimized with the CC3/aug-cc-pVTZ approach.
The equilibrium bond lengths
of \Cltwo{}, \Brtwo{}, and HBr were
taken from Ref.\ \onlinecite{herzberg4}.
In setting up and solving the DIP-EOMCC eigenvalue problems for the
\watertwoplus{}, \methanetwoplus{}, \bntwoplus{},
\Cltwoplus{}, \Brtwoplus{}, and \HBrtwoplus{}
target species and executing the preceding CC computations for
their neutral parents,
we used the restricted Hartree--Fock (RHF) orbitals of 
the respective \water{}, \methane{}, BN,
\Cltwo{}, \Brtwo{}, and HBr
molecules.
The relevant RHF reference determinants and transformed one- and two-electron integrals
were generated with the PySCF code,\cite{pyscf1,pyscf2} with which CCpy is interfaced.
The scalar-relativistic effects included in
our DIP-EOMCC calculations for the DIPs of \Cltwo{}, \Brtwo{}, and HBr
were handled using the SFX2C-1e spin-free exact two-component approach of Ref.\ \onlinecite{sfx2c1e}, as implemented
in PySCF, and the lowest-energy orbitals correlating with the chemical cores of
non-hydrogen atoms
were frozen in post-RHF steps.

The results of our DIP-EOMCCSD(3$h$-1$p$), DIP-EOMCCSD(4$h$-2$p$), DIP-EOMCCSD(T)(a)(4$h$-2$p$),
and DIP-EOMCCSDT(4$h$-2$p$) computations for the vertical DIPs of \water{}, \methane{}, and BN corresponding to the
lowest singlet and triplet states of the \watertwoplus{}, \methanetwoplus{}, and \bntwoplus{} dications are
summarized in Table \ref{table3}. The vertical DIPs of \Cltwo{}, \Brtwo{}, and HBr
corresponding to the ground and low-lying excited states of
\Cltwoplus{}, \Brtwoplus{}, and \HBrtwoplus{} resulting from
our DIP-EOMCCSD(3$h$-1$p$), DIP-EOMCCSD(4$h$-2$p$), DIP-EOMCCSD(T)(a)(4$h$-2$p$), and DIP-EOMCCSDT(4$h$-2$p$)
calculations are reported in Table \ref{table4}.
In much of the remaining discussion, we focus on the DIPs of \water{}, \methane{}, and BN shown in
Table \ref{table3}, where we compare our DIP-EOMCC data with their near-full-CI counterparts determined with the
help of CIPSI in Ref.\ \onlinecite{loos-ppbse}. While brief remarks about our DIP-EOMCC calculations for
the DIPs of \Cltwo{}, \Brtwo{}, and HBr summarized in Table \ref{table4} are given here as well, a more thorough
analysis of these computations is provided in the Supplementary Material.
As shown in
Table \ref{table3},
the vertical DIPs obtained in the 
DIP-EOMCCSD(3$h$-1$p$) calculations using the aug-cc-pVTZ basis set
are characterized by
significant errors relative to their near-exact counterparts
resulting from the CIPSI extrapolations performed in Ref.\ \onlinecite{loos-ppbse}, which are 0.77 and 0.61 eV
for the \watertriplet{} and \watersinglet{} states of \watertwoplus{},
0.32 and 0.31 eV for the \methanetriplet{} and \methanesinglet{} states of \methanetwoplus{},
and 0.44 and 0.35 eV for the \bntriplet{} and \bnsinglet{} states of \bntwoplus{}. 
For all of the calculated states of \watertwoplus{}, \methanetwoplus{}, and \bntwoplus{}
considered in Table \ref{table3} and for the majority of states of \Cltwoplus{}, \Brtwoplus{},
and \HBrtwoplus{} examined in Table \ref{table4}, especially when a larger cc-pVQZ basis set is employed,
the DIP-EOMCSCD(3$h$-1$p$) computations overestimate the corresponding DIPs of \water{}, \methane{}, BN,
\Cltwo{}, \Brtwo{}, and HBr relative to the reference data. In both sets of molecular examples
shown in Tables \ref{table3} and \ref{table4}, the poor quality of the DIPs obtained with the
DIP-EOMCCSD(3$h$-1$p$) approach
seems to be a consequence of neglecting the 4$h$-2$p$ component of $R_{\mu}^{(-2)}$. 

Indeed,
when $R_{\mu,4h\mbox{-}2p}$ is included
in $R_{\mu}^{(-2)}$ via the DIP-EOMCCSD(4$h$-2$p$) approach, the
errors in the lowest triplet and singlet DIPs of the water molecule
relative to the extrapolated CIPSI data obtained with DIP-EOMCCSD(3$h$-1$p$),
of 0.77 and 0.61 eV, reduce in the DIP-EOMCCSD(4$h$-2$p$) calculations to
0.25 and 0.25 eV, respectively. The analogous errors characterizing the
DIP-EOMCCSD(3$h$-1$p$) computations for methane, which are 0.32 and 0.31 eV,
reduce to 0.17 and 0.18 eV, respectively, when the DIP-EOMCCSD(4$h$-2$p$)
method is employed. Unfortunately,
the DIP-EOMCCSD(4$h$-2$p$) approach does not
always
improve the DIP-EOMCCSD(3$h$-1$p$) results.
For example,
the DIP values obtained in the 
DIP-EOMCCSD(4$h$-2$p$) calculations for BN
are less accurate than their
DIP-EOMCCSD(3$h$-1$p$)
counterparts, increasing the 
0.44 and 0.35 eV errors in the DIP-EOMCCSD(3$h$-1$p$) data
for the \bntriplet{} and \bnsinglet{} states of \bntwoplus{}
relative to their extrapolated CIPSI values to 0.61 and 0.67 eV, respectively,
when DIP-EOMCCSD(3$h$-1$p$) is replaced by DIP-EOMCCSD(4$h$-2$p$).
In general, the DIPs of \water{}, \methane{}, and BN
resulting from the DIP-EOMCCSD(4$h$-2$p$) calculations reported in 
Table \ref{table3}
lie substantially below the corresponding 
near-full-CI extrapolated CIPSI data. 
Similar remarks
apply to the DIPs of \Cltwo{}, \Brtwo{}, 
and HBr shown in Table \ref{table4}. This
behavior of DIP-EOMCCSD(4$h$-2$p$) can be attributed to the imbalance between the high-level
4$h$-2$p$ treatment of double ionization and
the lower-level
CCSD description of the neutral species.

The results of our DIP-EOMCCSDT(4$h$-2$p$) calculations, in which
the similarity-transformed Hamiltonian of CCSD is replaced by its CCSDT counterpart, allowing us
to treat the $N$- and ($N-2$)-electron species in a more accurate and balanced manner,
confirm 
this observation. As shown in Table \ref{table3}, the DIPs of \water{}, \methane{}, and BN
become much more accurate when we move from
the DIP-EOMCCSD(4$h$-2$p$) approach
to
its higher-level DIP-EOMCCSDT(4$h$-2$p$) counterpart. The 
errors in the DIPs associated with the lowest triplet and singlet dicationic states
of \water{}, \methane{}, and BN relative to the extrapolated CIPSI reference data, which are
0.25 and 0.25 eV for \watertwoplus{}, 0.17 and 0.18 eV for \methanetwoplus{}, and 
0.61 and 0.67 eV for \bntwoplus{}
in the
DIP-EOMCCSD(4$h$-2$p$) case,
reduce to 
the minuscule 0.02 and 0.03 eV for \watertwoplus{}, 0.00 and 0.01 eV for \methanetwoplus{},
and 0.01 and 0.00 eV for \bntwoplus{},
respectively, when the 
DIP-EOMCCSDT(4$h$-2$p$)
method is employed.
The same is generally true when examining the DIPs of \Cltwo{}, \Brtwo{},
and HBr shown in Table \ref{table4}, where the DIP-EOMCCSDT(4$h$-2$p$) approach offers
similar improvements. The only exceptions are the $c\:{^{1}}\Sigma_{u}^{-}$ state of
\Cltwoplus{} and the three states of \HBrtwoplus{}, for which the DIP-EOMCCSD(4$h$-2$p$)/cc-pVQZ
DIPs are already very accurate.
We conclude by
pointing out that the DIP-EOMCCSD(T)(a)(4$h$-2$p$) method, which offers
significant savings in the computational effort compared to full DIP-EOMCCSDT(4$h$-2$p$),
reproduces the DIPs of  
\water{}, \methane{}, and BN reported in Table \ref{table3} as well as the DIPs of
\Cltwo{}, \Brtwo{}, and HBr considered in Table \ref{table4} resulting from
the parent DIP-EOMCCSDT(4$h$-2$p$) calculations to within 0.02 eV. While we will continue testing
the DIP-EOMCCSD(T)(a)(4$h$-2$p$) approach against the DIP-EOMCCSDT(4$h$-2$p$) and other
high-accuracy data, its excellent performance in
this study
is encouraging.

In summary, we presented the fully factorized and programmable equations defining the DIP-EOMCCSDT(4$h$-2$p$)
approach and the perturbative approximation to it abbreviated as DIP-EOMCCSD(T)(a)(4$h$-2$p$). We incorporated
the resulting computer
codes
into
the open-source CCpy package available on GitHub. We applied the DIP-EOMCCSDT(4$h$-2$p$) and
DIP-EOMCCSD(T)(a)(4$h$-2$p$) methods and their DIP-EOMCCSD(3$h$-1$p$) and DIP-EOMCCSD(4$h$-2$p$) predecessors
to the vertical DIPs of 
\water{}, \methane{}, and BN, as described by the aug-cc-pVTZ basis set, and
\Cltwo{}, \Brtwo{}, and HBr,
using the cc-pVTZ and cc-pVQZ bases and
the spin-free two-component SFX2C-1e treatment of the scalar-relativistic effects. We demonstrated
that with the exception of the higher-lying $c\:{^{1}}\Sigma_{u}^{-}$ state of \Brtwoplus{},
the DIP values computed with
DIP-EOMCCSDT(4$h$-2$p$)
are not only in generally good agreement with the available
high-accuracy theoretical or experimental reference data,
but also more accurate than those
obtained with DIP-EOMCCSD(4$h$-2$p$),
which uses CCSD instead of CCSDT to construct
the underlying similarity-transformed Hamiltonian. The DIP-EOMCCSD(T)(a)(4$h$-2$p$) method, which avoids the
most expensive steps of DIP-EOMCCSDT(4$h$-2$p$), turned out to be similarly effective, recovering the vertical
DIPs of
\water{}, \methane{}, BN,
\Cltwo{}, \Brtwo{}, and HBr obtained with its DIP-EOMCCSDT(4$h$-2$p$) parent to within 0.02 eV.
Our future plans include the development of nonperturbative ways of
reducing costs of the DIP-EOMCCSDT(4$h$-2$p$) calculations through the
active-space treatments of CCSDT\cite{semi0b,semi2,semi4} and 4$h$-2$p$ amplitudes
\cite{jspp-dea-dip-2013,jspp-dea-dip-2014} and the use of
frozen natural orbitals, combined with Cholesky decomposition and density fitting techniques,
which will also be useful in improving our description of relativistic effects following
the four-component methodology of Ref.\ \onlinecite{4c-eomcc-fns}.

\section*{Supplementary Material}
See the Supplementary Material for the more detailed analysis of the DIPs of \Cltwo{}, \Brtwo{}, and HBr
resulting from the DIP-EOMCC calculations reported in Table \ref{table4}.

\begin{acknowledgments}
This work has been supported by the
Chemical Sciences, Geosciences and Biosciences
Division, Office of Basic Energy Sciences, Office of Science, 
U.S. Department of Energy
(Grant No.\ DE-FG02-01ER15228 to P.P). A.K.D. acknowledges support from SERB-India under
the CRG (Project No.\ CRG/2022/005672) and MATRICS (Project No.\ MTR/2021/000420) schemes.
We thank Dr. Jun Shen for inspecting Eqs.\ (\ref{eq2h})--(\ref{eq4h2p}).
\end{acknowledgments}

\section*{Data Availability}
The data that support the findings of this study are available within the article
and its supplementary material.


\section*{References}
\bibliographystyle{aipnum4-1}
\renewcommand{\baselinestretch}{1.1}
\bibliography{refs-revised-for-arxiv}

\begin{thebibliography}{75}%
\makeatletter
\providecommand \@ifxundefined [1]{%
 \@ifx{#1\undefined}
}%
\providecommand \@ifnum [1]{%
 \ifnum #1\expandafter \@firstoftwo
 \else \expandafter \@secondoftwo
 \fi
}%
\providecommand \@ifx [1]{%
 \ifx #1\expandafter \@firstoftwo
 \else \expandafter \@secondoftwo
 \fi
}%
\providecommand \natexlab [1]{#1}%
\providecommand \enquote  [1]{``#1''}%
\providecommand \bibnamefont  [1]{#1}%
\providecommand \bibfnamefont [1]{#1}%
\providecommand \citenamefont [1]{#1}%
\providecommand \href@noop [0]{\@secondoftwo}%
\providecommand \href [0]{\begingroup \@sanitize@url \@href}%
\providecommand \@href[1]{\@@startlink{#1}\@@href}%
\providecommand \@@href[1]{\endgroup#1\@@endlink}%
\providecommand \@sanitize@url [0]{\catcode `\\12\catcode `\$12\catcode
  `\&12\catcode `\#12\catcode `\^12\catcode `\_12\catcode `\%12\relax}%
\providecommand \@@startlink[1]{}%
\providecommand \@@endlink[0]{}%
\providecommand \url  [0]{\begingroup\@sanitize@url \@url }%
\providecommand \@url [1]{\endgroup\@href {#1}{\urlprefix }}%
\providecommand \urlprefix  [0]{URL }%
\providecommand \Eprint [0]{\href }%
\providecommand \doibase [0]{http://dx.doi.org/}%
\providecommand \selectlanguage [0]{\@gobble}%
\providecommand \bibinfo  [0]{\@secondoftwo}%
\providecommand \bibfield  [0]{\@secondoftwo}%
\providecommand \translation [1]{[#1]}%
\providecommand \BibitemOpen [0]{}%
\providecommand \bibitemStop [0]{}%
\providecommand \bibitemNoStop [0]{.\EOS\space}%
\providecommand \EOS [0]{\spacefactor3000\relax}%
\providecommand \BibitemShut  [1]{\csname bibitem#1\endcsname}%
\let\auto@bib@innerbib\@empty
\bibitem [{\citenamefont {Coester}(1958)}]{Coester:1958}%
  \BibitemOpen
  \bibfield  {author} {\bibinfo {author} {\bibfnamefont {F.}~\bibnamefont
  {Coester}},\ }\href@noop {} {\bibfield  {journal} {\bibinfo  {journal} {Nucl.
  Phys.}\ }\textbf {\bibinfo {volume} {7}},\ \bibinfo {pages} {421} (\bibinfo
  {year} {1958})}\BibitemShut {NoStop}%
\bibitem [{\citenamefont {Coester}\ and\ \citenamefont {K{\"
  u}mmel}(1960)}]{Coester:1960}%
  \BibitemOpen
  \bibfield  {author} {\bibinfo {author} {\bibfnamefont {F.}~\bibnamefont
  {Coester}}\ and\ \bibinfo {author} {\bibfnamefont {H.}~\bibnamefont {K{\"
  u}mmel}},\ }\href@noop {} {\bibfield  {journal} {\bibinfo  {journal} {Nucl.
  Phys.}\ }\textbf {\bibinfo {volume} {17}},\ \bibinfo {pages} {477} (\bibinfo
  {year} {1960})}\BibitemShut {NoStop}%
\bibitem [{\citenamefont {{\v C}{\'\i}{\v z}ek}(1966)}]{cizek1}%
  \BibitemOpen
  \bibfield  {author} {\bibinfo {author} {\bibfnamefont {J.}~\bibnamefont {{\v
  C}{\'\i}{\v z}ek}},\ }\href@noop {} {\bibfield  {journal} {\bibinfo
  {journal} {J. Chem. Phys.}\ }\textbf {\bibinfo {volume} {45}},\ \bibinfo
  {pages} {4256} (\bibinfo {year} {1966})}\BibitemShut {NoStop}%
\bibitem [{\citenamefont {{\v C}{\'\i}{\v z}ek}(1969)}]{cizek2}%
  \BibitemOpen
  \bibfield  {author} {\bibinfo {author} {\bibfnamefont {J.}~\bibnamefont {{\v
  C}{\'\i}{\v z}ek}},\ }\href@noop {} {\bibfield  {journal} {\bibinfo
  {journal} {Adv. Chem. Phys.}\ }\textbf {\bibinfo {volume} {14}},\ \bibinfo
  {pages} {35} (\bibinfo {year} {1969})}\BibitemShut {NoStop}%
\bibitem [{\citenamefont {Paldus}, \citenamefont {{\v C}{\'\i}{\v z}ek},\ and\
  \citenamefont {Shavitt}(1972)}]{cizek4}%
  \BibitemOpen
  \bibfield  {author} {\bibinfo {author} {\bibfnamefont {J.}~\bibnamefont
  {Paldus}}, \bibinfo {author} {\bibfnamefont {J.}~\bibnamefont {{\v
  C}{\'\i}{\v z}ek}}, \ and\ \bibinfo {author} {\bibfnamefont {I.}~\bibnamefont
  {Shavitt}},\ }\href@noop {} {\bibfield  {journal} {\bibinfo  {journal} {Phys.
  Rev. A}\ }\textbf {\bibinfo {volume} {5}},\ \bibinfo {pages} {50} (\bibinfo
  {year} {1972})}\BibitemShut {NoStop}%
\bibitem [{\citenamefont {Emrich}(1981{\natexlab{a}})}]{emrich}%
  \BibitemOpen
  \bibfield  {author} {\bibinfo {author} {\bibfnamefont {K.}~\bibnamefont
  {Emrich}},\ }\href@noop {} {\bibfield  {journal} {\bibinfo  {journal} {Nucl.
  Phys. A}\ }\textbf {\bibinfo {volume} {351}},\ \bibinfo {pages} {379}
  (\bibinfo {year} {1981}{\natexlab{a}})}\BibitemShut {NoStop}%
\bibitem [{\citenamefont {Emrich}(1981{\natexlab{b}})}]{emrich2}%
  \BibitemOpen
  \bibfield  {author} {\bibinfo {author} {\bibfnamefont {K.}~\bibnamefont
  {Emrich}},\ }\href@noop {} {\bibfield  {journal} {\bibinfo  {journal} {Nucl.
  Phys. A}\ }\textbf {\bibinfo {volume} {351}},\ \bibinfo {pages} {397}
  (\bibinfo {year} {1981}{\natexlab{b}})}\BibitemShut {NoStop}%
\bibitem [{\citenamefont {Geertsen}, \citenamefont {Rittby},\ and\
  \citenamefont {Bartlett}(1989)}]{eomcc1}%
  \BibitemOpen
  \bibfield  {author} {\bibinfo {author} {\bibfnamefont {J.}~\bibnamefont
  {Geertsen}}, \bibinfo {author} {\bibfnamefont {M.}~\bibnamefont {Rittby}}, \
  and\ \bibinfo {author} {\bibfnamefont {R.~J.}\ \bibnamefont {Bartlett}},\
  }\href@noop {} {\bibfield  {journal} {\bibinfo  {journal} {Chem. Phys.
  Lett.}\ }\textbf {\bibinfo {volume} {164}},\ \bibinfo {pages} {57} (\bibinfo
  {year} {1989})}\BibitemShut {NoStop}%
\bibitem [{\citenamefont {Comeau}\ and\ \citenamefont
  {Bartlett}(1993)}]{eomcc2}%
  \BibitemOpen
  \bibfield  {author} {\bibinfo {author} {\bibfnamefont {D.~C.}\ \bibnamefont
  {Comeau}}\ and\ \bibinfo {author} {\bibfnamefont {R.~J.}\ \bibnamefont
  {Bartlett}},\ }\href@noop {} {\bibfield  {journal} {\bibinfo  {journal}
  {Chem. Phys. Lett.}\ }\textbf {\bibinfo {volume} {207}},\ \bibinfo {pages}
  {414} (\bibinfo {year} {1993})}\BibitemShut {NoStop}%
\bibitem [{\citenamefont {Stanton}\ and\ \citenamefont
  {Bartlett}(1993)}]{eomcc3}%
  \BibitemOpen
  \bibfield  {author} {\bibinfo {author} {\bibfnamefont {J.~F.}\ \bibnamefont
  {Stanton}}\ and\ \bibinfo {author} {\bibfnamefont {R.~J.}\ \bibnamefont
  {Bartlett}},\ }\href@noop {} {\bibfield  {journal} {\bibinfo  {journal} {J.
  Chem. Phys.}\ }\textbf {\bibinfo {volume} {98}},\ \bibinfo {pages} {7029}
  (\bibinfo {year} {1993})}\BibitemShut {NoStop}%
\bibitem [{\citenamefont {Nooijen}\ and\ \citenamefont
  {Bartlett}(1995{\natexlab{a}})}]{eaccsd1}%
  \BibitemOpen
  \bibfield  {author} {\bibinfo {author} {\bibfnamefont {M.}~\bibnamefont
  {Nooijen}}\ and\ \bibinfo {author} {\bibfnamefont {R.~J.}\ \bibnamefont
  {Bartlett}},\ }\href@noop {} {\bibfield  {journal} {\bibinfo  {journal} {J.
  Chem. Phys.}\ }\textbf {\bibinfo {volume} {102}},\ \bibinfo {pages} {3629}
  (\bibinfo {year} {1995}{\natexlab{a}})}\BibitemShut {NoStop}%
\bibitem [{\citenamefont {Nooijen}\ and\ \citenamefont
  {Bartlett}(1995{\natexlab{b}})}]{eaccsd2}%
  \BibitemOpen
  \bibfield  {author} {\bibinfo {author} {\bibfnamefont {M.}~\bibnamefont
  {Nooijen}}\ and\ \bibinfo {author} {\bibfnamefont {R.~J.}\ \bibnamefont
  {Bartlett}},\ }\href@noop {} {\bibfield  {journal} {\bibinfo  {journal} {J.
  Chem. Phys.}\ }\textbf {\bibinfo {volume} {102}},\ \bibinfo {pages} {6735}
  (\bibinfo {year} {1995}{\natexlab{b}})}\BibitemShut {NoStop}%
\bibitem [{\citenamefont {Hirata}, \citenamefont {Nooijen},\ and\ \citenamefont
  {Bartlett}(2000)}]{hirataeaip}%
  \BibitemOpen
  \bibfield  {author} {\bibinfo {author} {\bibfnamefont {S.}~\bibnamefont
  {Hirata}}, \bibinfo {author} {\bibfnamefont {M.}~\bibnamefont {Nooijen}}, \
  and\ \bibinfo {author} {\bibfnamefont {R.~J.}\ \bibnamefont {Bartlett}},\
  }\href@noop {} {\bibfield  {journal} {\bibinfo  {journal} {Chem. Phys.
  Lett.}\ }\textbf {\bibinfo {volume} {328}},\ \bibinfo {pages} {459} (\bibinfo
  {year} {2000})}\BibitemShut {NoStop}%
\bibitem [{\citenamefont {Musia{\l}}\ and\ \citenamefont
  {Bartlett}(2003)}]{eaccsdt1}%
  \BibitemOpen
  \bibfield  {author} {\bibinfo {author} {\bibfnamefont {M.}~\bibnamefont
  {Musia{\l}}}\ and\ \bibinfo {author} {\bibfnamefont {R.~J.}\ \bibnamefont
  {Bartlett}},\ }\href@noop {} {\bibfield  {journal} {\bibinfo  {journal} {J.
  Chem. Phys.}\ }\textbf {\bibinfo {volume} {119}},\ \bibinfo {pages} {1901}
  (\bibinfo {year} {2003})}\BibitemShut {NoStop}%
\bibitem [{\citenamefont {Gour}, \citenamefont {Piecuch},\ and\ \citenamefont
  {W{\l}och}(2005)}]{gour1}%
  \BibitemOpen
  \bibfield  {author} {\bibinfo {author} {\bibfnamefont {J.~R.}\ \bibnamefont
  {Gour}}, \bibinfo {author} {\bibfnamefont {P.}~\bibnamefont {Piecuch}}, \
  and\ \bibinfo {author} {\bibfnamefont {M.}~\bibnamefont {W{\l}och}},\
  }\href@noop {} {\bibfield  {journal} {\bibinfo  {journal} {J. Chem. Phys.}\
  }\textbf {\bibinfo {volume} {123}},\ \bibinfo {pages} {134113} (\bibinfo
  {year} {2005})}\BibitemShut {NoStop}%
\bibitem [{\citenamefont {Gour}, \citenamefont {Piecuch},\ and\ \citenamefont
  {W{\l}och}(2006)}]{gour2}%
  \BibitemOpen
  \bibfield  {author} {\bibinfo {author} {\bibfnamefont {J.~R.}\ \bibnamefont
  {Gour}}, \bibinfo {author} {\bibfnamefont {P.}~\bibnamefont {Piecuch}}, \
  and\ \bibinfo {author} {\bibfnamefont {M.}~\bibnamefont {W{\l}och}},\
  }\href@noop {} {\bibfield  {journal} {\bibinfo  {journal} {Int. J. Quantum
  Chem.}\ }\textbf {\bibinfo {volume} {106}},\ \bibinfo {pages} {2854}
  (\bibinfo {year} {2006})}\BibitemShut {NoStop}%
\bibitem [{\citenamefont {Gour}\ and\ \citenamefont {Piecuch}(2006)}]{gour3}%
  \BibitemOpen
  \bibfield  {author} {\bibinfo {author} {\bibfnamefont {J.~R.}\ \bibnamefont
  {Gour}}\ and\ \bibinfo {author} {\bibfnamefont {P.}~\bibnamefont {Piecuch}},\
  }\href@noop {} {\bibfield  {journal} {\bibinfo  {journal} {J. Chem. Phys.}\
  }\textbf {\bibinfo {volume} {125}},\ \bibinfo {pages} {234107} (\bibinfo
  {year} {2006})}\BibitemShut {NoStop}%
\bibitem [{\citenamefont {Nooijen}\ and\ \citenamefont
  {Snijders}(1992)}]{ipccsd2}%
  \BibitemOpen
  \bibfield  {author} {\bibinfo {author} {\bibfnamefont {M.}~\bibnamefont
  {Nooijen}}\ and\ \bibinfo {author} {\bibfnamefont {J.~G.}\ \bibnamefont
  {Snijders}},\ }\href@noop {} {\bibfield  {journal} {\bibinfo  {journal} {Int.
  J. Quantum Chem. Symp.}\ }\textbf {\bibinfo {volume} {26}},\ \bibinfo {pages}
  {55} (\bibinfo {year} {1992})}\BibitemShut {NoStop}%
\bibitem [{\citenamefont {Nooijen}\ and\ \citenamefont
  {Snijders}(1993)}]{ipccsd3}%
  \BibitemOpen
  \bibfield  {author} {\bibinfo {author} {\bibfnamefont {M.}~\bibnamefont
  {Nooijen}}\ and\ \bibinfo {author} {\bibfnamefont {J.~G.}\ \bibnamefont
  {Snijders}},\ }\href@noop {} {\bibfield  {journal} {\bibinfo  {journal} {Int.
  J. Quantum Chem.}\ }\textbf {\bibinfo {volume} {48}},\ \bibinfo {pages} {15}
  (\bibinfo {year} {1993})}\BibitemShut {NoStop}%
\bibitem [{\citenamefont {Stanton}\ and\ \citenamefont
  {Gauss}(1994)}]{ipccsd4}%
  \BibitemOpen
  \bibfield  {author} {\bibinfo {author} {\bibfnamefont {J.~F.}\ \bibnamefont
  {Stanton}}\ and\ \bibinfo {author} {\bibfnamefont {J.}~\bibnamefont
  {Gauss}},\ }\href@noop {} {\bibfield  {journal} {\bibinfo  {journal} {J.
  Chem. Phys.}\ }\textbf {\bibinfo {volume} {101}},\ \bibinfo {pages} {8938}
  (\bibinfo {year} {1994})}\BibitemShut {NoStop}%
\bibitem [{\citenamefont {Bartlett}\ and\ \citenamefont
  {Stanton}(1994)}]{ipccsd1}%
  \BibitemOpen
  \bibfield  {author} {\bibinfo {author} {\bibfnamefont {R.~J.}\ \bibnamefont
  {Bartlett}}\ and\ \bibinfo {author} {\bibfnamefont {J.~F.}\ \bibnamefont
  {Stanton}},\ }in\ \href@noop {} {\emph {\bibinfo {booktitle} {Reviews in
  Computational Chemistry}}},\ Vol.~\bibinfo {volume} {5},\ \bibinfo {editor}
  {edited by\ \bibinfo {editor} {\bibfnamefont {K.~B.}\ \bibnamefont
  {Lipkowitz}}\ and\ \bibinfo {editor} {\bibfnamefont {D.~B.}\ \bibnamefont
  {Boyd}}}\ (\bibinfo  {publisher} {VCH Publishers},\ \bibinfo {address} {New
  York},\ \bibinfo {year} {1994})\ pp.\ \bibinfo {pages} {65--169}\BibitemShut
  {NoStop}%
\bibitem [{\citenamefont {Musia{\l}}, \citenamefont {Kucharski},\ and\
  \citenamefont {Bartlett}(2003)}]{ipccsdt1}%
  \BibitemOpen
  \bibfield  {author} {\bibinfo {author} {\bibfnamefont {M.}~\bibnamefont
  {Musia{\l}}}, \bibinfo {author} {\bibfnamefont {S.~A.}\ \bibnamefont
  {Kucharski}}, \ and\ \bibinfo {author} {\bibfnamefont {R.~J.}\ \bibnamefont
  {Bartlett}},\ }\href@noop {} {\bibfield  {journal} {\bibinfo  {journal} {J.
  Chem. Phys.}\ }\textbf {\bibinfo {volume} {118}},\ \bibinfo {pages} {1128}
  (\bibinfo {year} {2003})}\BibitemShut {NoStop}%
\bibitem [{\citenamefont {Musia{\l}}\ and\ \citenamefont
  {Bartlett}(2004)}]{ipccsdt2}%
  \BibitemOpen
  \bibfield  {author} {\bibinfo {author} {\bibfnamefont {M.}~\bibnamefont
  {Musia{\l}}}\ and\ \bibinfo {author} {\bibfnamefont {R.~J.}\ \bibnamefont
  {Bartlett}},\ }\href@noop {} {\bibfield  {journal} {\bibinfo  {journal}
  {Chem. Phys. Lett.}\ }\textbf {\bibinfo {volume} {384}},\ \bibinfo {pages}
  {210} (\bibinfo {year} {2004})}\BibitemShut {NoStop}%
\bibitem [{\citenamefont {Bomble}\ \emph {et~al.}(2005)\citenamefont {Bomble},
  \citenamefont {Saeh}, \citenamefont {Stanton}, \citenamefont {Szalay},
  \citenamefont {K{\' a}llay},\ and\ \citenamefont {Gauss}}]{ipccsdt-3}%
  \BibitemOpen
  \bibfield  {author} {\bibinfo {author} {\bibfnamefont {Y.~J.}\ \bibnamefont
  {Bomble}}, \bibinfo {author} {\bibfnamefont {J.~C.}\ \bibnamefont {Saeh}},
  \bibinfo {author} {\bibfnamefont {J.~F.}\ \bibnamefont {Stanton}}, \bibinfo
  {author} {\bibfnamefont {P.~G.}\ \bibnamefont {Szalay}}, \bibinfo {author}
  {\bibfnamefont {M.}~\bibnamefont {K{\' a}llay}}, \ and\ \bibinfo {author}
  {\bibfnamefont {J.}~\bibnamefont {Gauss}},\ }\href@noop {} {\bibfield
  {journal} {\bibinfo  {journal} {J. Chem. Phys.}\ }\textbf {\bibinfo {volume}
  {122}},\ \bibinfo {pages} {154107} (\bibinfo {year} {2005})}\BibitemShut
  {NoStop}%
\bibitem [{\citenamefont {Kamiya}\ and\ \citenamefont
  {Hirata}(2006)}]{hirata_ea_ip}%
  \BibitemOpen
  \bibfield  {author} {\bibinfo {author} {\bibfnamefont {M.}~\bibnamefont
  {Kamiya}}\ and\ \bibinfo {author} {\bibfnamefont {S.}~\bibnamefont
  {Hirata}},\ }\href@noop {} {\bibfield  {journal} {\bibinfo  {journal} {J.
  Chem. Phys.}\ }\textbf {\bibinfo {volume} {125}},\ \bibinfo {pages} {074111}
  (\bibinfo {year} {2006})}\BibitemShut {NoStop}%
\bibitem [{\citenamefont {Nooijen}\ and\ \citenamefont
  {Bartlett}(1997{\natexlab{a}})}]{dipea1}%
  \BibitemOpen
  \bibfield  {author} {\bibinfo {author} {\bibfnamefont {M.}~\bibnamefont
  {Nooijen}}\ and\ \bibinfo {author} {\bibfnamefont {R.~J.}\ \bibnamefont
  {Bartlett}},\ }\href@noop {} {\bibfield  {journal} {\bibinfo  {journal} {J.
  Chem. Phys.}\ }\textbf {\bibinfo {volume} {106}},\ \bibinfo {pages} {6441}
  (\bibinfo {year} {1997}{\natexlab{a}})}\BibitemShut {NoStop}%
\bibitem [{\citenamefont {Nooijen}\ and\ \citenamefont
  {Bartlett}(1997{\natexlab{b}})}]{steom2}%
  \BibitemOpen
  \bibfield  {author} {\bibinfo {author} {\bibfnamefont {M.}~\bibnamefont
  {Nooijen}}\ and\ \bibinfo {author} {\bibfnamefont {R.~J.}\ \bibnamefont
  {Bartlett}},\ }\href@noop {} {\bibfield  {journal} {\bibinfo  {journal} {J.
  Chem. Phys.}\ }\textbf {\bibinfo {volume} {107}},\ \bibinfo {pages} {6812}
  (\bibinfo {year} {1997}{\natexlab{b}})}\BibitemShut {NoStop}%
\bibitem [{\citenamefont {Wladyslawski}\ and\ \citenamefont
  {Nooijen}(2002)}]{dipea2}%
  \BibitemOpen
  \bibfield  {author} {\bibinfo {author} {\bibfnamefont {M.}~\bibnamefont
  {Wladyslawski}}\ and\ \bibinfo {author} {\bibfnamefont {M.}~\bibnamefont
  {Nooijen}},\ }in\ \href@noop {} {\emph {\bibinfo {booktitle} {Low-Lying
  Potential Energy Surfaces}}},\ \bibinfo {series} {ACS Symposium Series},
  Vol.\ \bibinfo {volume} {828},\ \bibinfo {editor} {edited by\ \bibinfo
  {editor} {\bibfnamefont {M.~R.}\ \bibnamefont {Hoffmann}}\ and\ \bibinfo
  {editor} {\bibfnamefont {K.~G.}\ \bibnamefont {Dyall}}}\ (\bibinfo
  {publisher} {American Chemical Society},\ \bibinfo {address} {{Washington,
  D.C.}},\ \bibinfo {year} {2002})\ pp.\ \bibinfo {pages} {65--92}\BibitemShut
  {NoStop}%
\bibitem [{\citenamefont {Nooijen}(2002)}]{dipea3}%
  \BibitemOpen
  \bibfield  {author} {\bibinfo {author} {\bibfnamefont {M.}~\bibnamefont
  {Nooijen}},\ }\href@noop {} {\bibfield  {journal} {\bibinfo  {journal} {Int.
  J. Mol. Sci.}\ }\textbf {\bibinfo {volume} {3}},\ \bibinfo {pages} {656}
  (\bibinfo {year} {2002})}\BibitemShut {NoStop}%
\bibitem [{\citenamefont {Sattelmeyer}, \citenamefont {Schaefer},\ and\
  \citenamefont {Stanton}(2003)}]{dip-stanton}%
  \BibitemOpen
  \bibfield  {author} {\bibinfo {author} {\bibfnamefont {K.~W.}\ \bibnamefont
  {Sattelmeyer}}, \bibinfo {author} {\bibfnamefont {H.~F.}\ \bibnamefont
  {Schaefer}, \bibfnamefont {III}}, \ and\ \bibinfo {author} {\bibfnamefont
  {J.~F.}\ \bibnamefont {Stanton}},\ }\href@noop {} {\bibfield  {journal}
  {\bibinfo  {journal} {Chem. Phys. Lett.}\ }\textbf {\bibinfo {volume}
  {378}},\ \bibinfo {pages} {42} (\bibinfo {year} {2003})}\BibitemShut
  {NoStop}%
\bibitem [{\citenamefont {Musia{\l}}, \citenamefont {Perera},\ and\
  \citenamefont {Bartlett}(2011)}]{dipea5}%
  \BibitemOpen
  \bibfield  {author} {\bibinfo {author} {\bibfnamefont {M.}~\bibnamefont
  {Musia{\l}}}, \bibinfo {author} {\bibfnamefont {A.}~\bibnamefont {Perera}}, \
  and\ \bibinfo {author} {\bibfnamefont {R.~J.}\ \bibnamefont {Bartlett}},\
  }\href@noop {} {\bibfield  {journal} {\bibinfo  {journal} {J. Chem. Phys.}\
  }\textbf {\bibinfo {volume} {134}},\ \bibinfo {pages} {114108} (\bibinfo
  {year} {2011})}\BibitemShut {NoStop}%
\bibitem [{\citenamefont {Musia{\l}}, \citenamefont {Kucharski},\ and\
  \citenamefont {Bartlett}(2011)}]{dipea6}%
  \BibitemOpen
  \bibfield  {author} {\bibinfo {author} {\bibfnamefont {M.}~\bibnamefont
  {Musia{\l}}}, \bibinfo {author} {\bibfnamefont {S.~A.}\ \bibnamefont
  {Kucharski}}, \ and\ \bibinfo {author} {\bibfnamefont {R.~J.}\ \bibnamefont
  {Bartlett}},\ }\href@noop {} {\bibfield  {journal} {\bibinfo  {journal} {J.
  Chem. Theory Comput.}\ }\textbf {\bibinfo {volume} {7}},\ \bibinfo {pages}
  {3088} (\bibinfo {year} {2011})}\BibitemShut {NoStop}%
\bibitem [{\citenamefont {Ku{\' s}}\ and\ \citenamefont
  {Krylov}(2011)}]{kus-krylov-2011}%
  \BibitemOpen
  \bibfield  {author} {\bibinfo {author} {\bibfnamefont {T.}~\bibnamefont
  {Ku{\' s}}}\ and\ \bibinfo {author} {\bibfnamefont {A.~I.}\ \bibnamefont
  {Krylov}},\ }\href@noop {} {\bibfield  {journal} {\bibinfo  {journal} {J.
  Chem. Phys.}\ }\textbf {\bibinfo {volume} {135}},\ \bibinfo {pages} {084109}
  (\bibinfo {year} {2011})}\BibitemShut {NoStop}%
\bibitem [{\citenamefont {Ku{\' s}}\ and\ \citenamefont
  {Krylov}(2012)}]{kus-krylov-2012}%
  \BibitemOpen
  \bibfield  {author} {\bibinfo {author} {\bibfnamefont {T.}~\bibnamefont
  {Ku{\' s}}}\ and\ \bibinfo {author} {\bibfnamefont {A.~I.}\ \bibnamefont
  {Krylov}},\ }\href@noop {} {\bibfield  {journal} {\bibinfo  {journal} {J.
  Chem. Phys.}\ }\textbf {\bibinfo {volume} {136}},\ \bibinfo {pages} {244109}
  (\bibinfo {year} {2012})}\BibitemShut {NoStop}%
\bibitem [{\citenamefont {Shen}\ and\ \citenamefont
  {Piecuch}(2013)}]{jspp-dea-dip-2013}%
  \BibitemOpen
  \bibfield  {author} {\bibinfo {author} {\bibfnamefont {J.}~\bibnamefont
  {Shen}}\ and\ \bibinfo {author} {\bibfnamefont {P.}~\bibnamefont {Piecuch}},\
  }\href@noop {} {\bibfield  {journal} {\bibinfo  {journal} {J. Chem. Phys.}\
  }\textbf {\bibinfo {volume} {138}},\ \bibinfo {pages} {194102} (\bibinfo
  {year} {2013})}\BibitemShut {NoStop}%
\bibitem [{\citenamefont {Shen}\ and\ \citenamefont
  {Piecuch}(2014)}]{jspp-dea-dip-2014}%
  \BibitemOpen
  \bibfield  {author} {\bibinfo {author} {\bibfnamefont {J.}~\bibnamefont
  {Shen}}\ and\ \bibinfo {author} {\bibfnamefont {P.}~\bibnamefont {Piecuch}},\
  }\href@noop {} {\bibfield  {journal} {\bibinfo  {journal} {Mol. Phys.}\
  }\textbf {\bibinfo {volume} {112}},\ \bibinfo {pages} {868} (\bibinfo {year}
  {2014})}\BibitemShut {NoStop}%
\bibitem [{\citenamefont {Ajala}, \citenamefont {Shen},\ and\ \citenamefont
  {Piecuch}(2017)}]{jspp-dea-dip-2017}%
  \BibitemOpen
  \bibfield  {author} {\bibinfo {author} {\bibfnamefont {A.~O.}\ \bibnamefont
  {Ajala}}, \bibinfo {author} {\bibfnamefont {J.}~\bibnamefont {Shen}}, \ and\
  \bibinfo {author} {\bibfnamefont {P.}~\bibnamefont {Piecuch}},\ }\href@noop
  {} {\bibfield  {journal} {\bibinfo  {journal} {J. Phys. Chem. A}\ }\textbf
  {\bibinfo {volume} {121}},\ \bibinfo {pages} {3469} (\bibinfo {year}
  {2017})}\BibitemShut {NoStop}%
\bibitem [{\citenamefont {Shen}\ and\ \citenamefont
  {Piecuch}(2021)}]{jspp-dea-dip-2021}%
  \BibitemOpen
  \bibfield  {author} {\bibinfo {author} {\bibfnamefont {J.}~\bibnamefont
  {Shen}}\ and\ \bibinfo {author} {\bibfnamefont {P.}~\bibnamefont {Piecuch}},\
  }\href@noop {} {\bibfield  {journal} {\bibinfo  {journal} {Mol. Phys.}\
  }\textbf {\bibinfo {volume} {119}},\ \bibinfo {pages} {e1966534} (\bibinfo
  {year} {2021})}\BibitemShut {NoStop}%
\bibitem [{\citenamefont {Gulania}\ \emph {et~al.}(2021)\citenamefont
  {Gulania}, \citenamefont {Kj$\phi$nstad}, \citenamefont {Stanton},
  \citenamefont {Koch},\ and\ \citenamefont {Krylov}}]{dipea7}%
  \BibitemOpen
  \bibfield  {author} {\bibinfo {author} {\bibfnamefont {S.}~\bibnamefont
  {Gulania}}, \bibinfo {author} {\bibfnamefont {E.~F.}\ \bibnamefont
  {Kj$\phi$nstad}}, \bibinfo {author} {\bibfnamefont {J.~F.}\ \bibnamefont
  {Stanton}}, \bibinfo {author} {\bibfnamefont {H.}~\bibnamefont {Koch}}, \
  and\ \bibinfo {author} {\bibfnamefont {A.~I.}\ \bibnamefont {Krylov}},\
  }\href@noop {} {\bibfield  {journal} {\bibinfo  {journal} {J. Chem. Phys.}\
  }\textbf {\bibinfo {volume} {154}},\ \bibinfo {pages} {114115} (\bibinfo
  {year} {2021})}\BibitemShut {NoStop}%
\bibitem [{\citenamefont {Musia{\l}}\ \emph {et~al.}(2012)\citenamefont
  {Musia{\l}}, \citenamefont {Olsz{\' o}wka}, \citenamefont {Lyakh},\ and\
  \citenamefont {Bartlett}}]{tea-eomcc}%
  \BibitemOpen
  \bibfield  {author} {\bibinfo {author} {\bibfnamefont {M.}~\bibnamefont
  {Musia{\l}}}, \bibinfo {author} {\bibfnamefont {M.}~\bibnamefont {Olsz{\'
  o}wka}}, \bibinfo {author} {\bibfnamefont {D.~I.}\ \bibnamefont {Lyakh}}, \
  and\ \bibinfo {author} {\bibfnamefont {R.~J.}\ \bibnamefont {Bartlett}},\
  }\href@noop {} {\bibfield  {journal} {\bibinfo  {journal} {J. Chem. Phys.}\
  }\textbf {\bibinfo {volume} {137}},\ \bibinfo {pages} {174102} (\bibinfo
  {year} {2012})}\BibitemShut {NoStop}%
\bibitem [{\citenamefont {Ghosh}, \citenamefont {Vaval},\ and\ \citenamefont
  {Pal}(2017)}]{ghosh-auger}%
  \BibitemOpen
  \bibfield  {author} {\bibinfo {author} {\bibfnamefont {A.}~\bibnamefont
  {Ghosh}}, \bibinfo {author} {\bibfnamefont {N.}~\bibnamefont {Vaval}}, \ and\
  \bibinfo {author} {\bibfnamefont {S.}~\bibnamefont {Pal}},\ }\href@noop {}
  {\bibfield  {journal} {\bibinfo  {journal} {Chem. Phys.}\ }\textbf {\bibinfo
  {volume} {482}},\ \bibinfo {pages} {160} (\bibinfo {year}
  {2017})}\BibitemShut {NoStop}%
\bibitem [{\citenamefont {Skomorowski}\ and\ \citenamefont
  {Krylov}(2021{\natexlab{a}})}]{krylov-auger-1}%
  \BibitemOpen
  \bibfield  {author} {\bibinfo {author} {\bibfnamefont {W.}~\bibnamefont
  {Skomorowski}}\ and\ \bibinfo {author} {\bibfnamefont {A.~I.}\ \bibnamefont
  {Krylov}},\ }\href@noop {} {\bibfield  {journal} {\bibinfo  {journal} {J.
  Chem. Phys.}\ }\textbf {\bibinfo {volume} {154}},\ \bibinfo {pages} {084124}
  (\bibinfo {year} {2021}{\natexlab{a}})}\BibitemShut {NoStop}%
\bibitem [{\citenamefont {Skomorowski}\ and\ \citenamefont
  {Krylov}(2021{\natexlab{b}})}]{krylov-auger-2}%
  \BibitemOpen
  \bibfield  {author} {\bibinfo {author} {\bibfnamefont {W.}~\bibnamefont
  {Skomorowski}}\ and\ \bibinfo {author} {\bibfnamefont {A.~I.}\ \bibnamefont
  {Krylov}},\ }\href@noop {} {\bibfield  {journal} {\bibinfo  {journal} {J.
  Chem. Phys.}\ }\textbf {\bibinfo {volume} {154}},\ \bibinfo {pages} {084125}
  (\bibinfo {year} {2021}{\natexlab{b}})}\BibitemShut {NoStop}%
\bibitem [{\citenamefont {Jayadev}\ \emph {et~al.}(2023)\citenamefont
  {Jayadev}, \citenamefont {Ferino-P{\'e}rez}, \citenamefont {Matz},
  \citenamefont {Krylov},\ and\ \citenamefont {Jagau}}]{auger-benzene}%
  \BibitemOpen
  \bibfield  {author} {\bibinfo {author} {\bibfnamefont {N.~K.}\ \bibnamefont
  {Jayadev}}, \bibinfo {author} {\bibfnamefont {A.}~\bibnamefont
  {Ferino-P{\'e}rez}}, \bibinfo {author} {\bibfnamefont {F.}~\bibnamefont
  {Matz}}, \bibinfo {author} {\bibfnamefont {A.~I.}\ \bibnamefont {Krylov}}, \
  and\ \bibinfo {author} {\bibfnamefont {T.-C.}\ \bibnamefont {Jagau}},\
  }\href@noop {} {\bibfield  {journal} {\bibinfo  {journal} {J. Chem. Phys.}\
  }\textbf {\bibinfo {volume} {158}},\ \bibinfo {pages} {064109} (\bibinfo
  {year} {2023})}\BibitemShut {NoStop}%
\bibitem [{\citenamefont {Stamm}\ \emph {et~al.}(2025)\citenamefont {Stamm},
  \citenamefont {Priyadarsini}, \citenamefont {Sandhu}, \citenamefont
  {Chakraborty}, \citenamefont {Shen}, \citenamefont {Kwon}, \citenamefont
  {Sandhu}, \citenamefont {Wicka}, \citenamefont {Mehmood}, \citenamefont
  {Levine}, \citenamefont {Piecuch},\ and\ \citenamefont
  {Dantus}}]{nat-commun-h3plus}%
  \BibitemOpen
  \bibfield  {author} {\bibinfo {author} {\bibfnamefont {J.}~\bibnamefont
  {Stamm}}, \bibinfo {author} {\bibfnamefont {S.~S.}\ \bibnamefont
  {Priyadarsini}}, \bibinfo {author} {\bibfnamefont {S.}~\bibnamefont
  {Sandhu}}, \bibinfo {author} {\bibfnamefont {A.}~\bibnamefont {Chakraborty}},
  \bibinfo {author} {\bibfnamefont {J.}~\bibnamefont {Shen}}, \bibinfo {author}
  {\bibfnamefont {S.}~\bibnamefont {Kwon}}, \bibinfo {author} {\bibfnamefont
  {J.}~\bibnamefont {Sandhu}}, \bibinfo {author} {\bibfnamefont
  {C.}~\bibnamefont {Wicka}}, \bibinfo {author} {\bibfnamefont
  {A.}~\bibnamefont {Mehmood}}, \bibinfo {author} {\bibfnamefont {B.~G.}\
  \bibnamefont {Levine}}, \bibinfo {author} {\bibfnamefont {P.}~\bibnamefont
  {Piecuch}}, \ and\ \bibinfo {author} {\bibfnamefont {M.}~\bibnamefont
  {Dantus}},\ }\href@noop {} {\bibfield  {journal} {\bibinfo  {journal} {Nat.
  Commun.}\ }\textbf {\bibinfo {volume} {16}},\ \bibinfo {pages} {410}
  (\bibinfo {year} {2025})}\BibitemShut {NoStop}%
\bibitem [{\citenamefont {Purvis}\ and\ \citenamefont {Bartlett}(1982)}]{ccsd}%
  \BibitemOpen
  \bibfield  {author} {\bibinfo {author} {\bibfnamefont {G.~D.}\ \bibnamefont
  {Purvis}, \bibfnamefont {III}}\ and\ \bibinfo {author} {\bibfnamefont
  {R.~J.}\ \bibnamefont {Bartlett}},\ }\href@noop {} {\bibfield  {journal}
  {\bibinfo  {journal} {J. Chem. Phys.}\ }\textbf {\bibinfo {volume} {76}},\
  \bibinfo {pages} {1910} (\bibinfo {year} {1982})}\BibitemShut {NoStop}%
\bibitem [{\citenamefont {Cullen}\ and\ \citenamefont {Zerner}(1982)}]{ccsd2}%
  \BibitemOpen
  \bibfield  {author} {\bibinfo {author} {\bibfnamefont {J.~M.}\ \bibnamefont
  {Cullen}}\ and\ \bibinfo {author} {\bibfnamefont {M.~C.}\ \bibnamefont
  {Zerner}},\ }\href@noop {} {\bibfield  {journal} {\bibinfo  {journal} {J.
  Chem. Phys.}\ }\textbf {\bibinfo {volume} {77}},\ \bibinfo {pages} {4088}
  (\bibinfo {year} {1982})}\BibitemShut {NoStop}%
\bibitem [{\citenamefont {Scuseria}\ \emph {et~al.}(1987)\citenamefont
  {Scuseria}, \citenamefont {Scheiner}, \citenamefont {Lee}, \citenamefont
  {Rice},\ and\ \citenamefont {Schaefer}}]{ccsdfritz}%
  \BibitemOpen
  \bibfield  {author} {\bibinfo {author} {\bibfnamefont {G.~E.}\ \bibnamefont
  {Scuseria}}, \bibinfo {author} {\bibfnamefont {A.~C.}\ \bibnamefont
  {Scheiner}}, \bibinfo {author} {\bibfnamefont {T.~J.}\ \bibnamefont {Lee}},
  \bibinfo {author} {\bibfnamefont {J.~E.}\ \bibnamefont {Rice}}, \ and\
  \bibinfo {author} {\bibfnamefont {H.~F.}\ \bibnamefont {Schaefer},
  \bibfnamefont {III}},\ }\href@noop {} {\bibfield  {journal} {\bibinfo
  {journal} {J. Chem. Phys.}\ }\textbf {\bibinfo {volume} {86}},\ \bibinfo
  {pages} {2881} (\bibinfo {year} {1987})}\BibitemShut {NoStop}%
\bibitem [{\citenamefont {Piecuch}\ and\ \citenamefont
  {Paldus}(1989)}]{osaccsd}%
  \BibitemOpen
  \bibfield  {author} {\bibinfo {author} {\bibfnamefont {P.}~\bibnamefont
  {Piecuch}}\ and\ \bibinfo {author} {\bibfnamefont {J.}~\bibnamefont
  {Paldus}},\ }\href@noop {} {\bibfield  {journal} {\bibinfo  {journal} {Int.
  J. Quantum Chem.}\ }\textbf {\bibinfo {volume} {36}},\ \bibinfo {pages} {429}
  (\bibinfo {year} {1989})}\BibitemShut {NoStop}%
\bibitem [{\citenamefont {Noga}\ and\ \citenamefont
  {Bartlett}(1987)}]{ccfullt}%
  \BibitemOpen
  \bibfield  {author} {\bibinfo {author} {\bibfnamefont {J.}~\bibnamefont
  {Noga}}\ and\ \bibinfo {author} {\bibfnamefont {R.~J.}\ \bibnamefont
  {Bartlett}},\ }\href@noop {} {\bibfield  {journal} {\bibinfo  {journal} {J.
  Chem. Phys.}\ }\textbf {\bibinfo {volume} {86}},\ \bibinfo {pages} {7041}
  (\bibinfo {year} {1987})},\ \bibinfo {note} {{\bf 89}, 3401 (1988)
  [Erratum]}\BibitemShut {NoStop}%
\bibitem [{\citenamefont {Scuseria}\ and\ \citenamefont
  {Schaefer}(1988)}]{ccfullt2}%
  \BibitemOpen
  \bibfield  {author} {\bibinfo {author} {\bibfnamefont {G.~E.}\ \bibnamefont
  {Scuseria}}\ and\ \bibinfo {author} {\bibfnamefont {H.~F.}\ \bibnamefont
  {Schaefer}, \bibfnamefont {III}},\ }\href@noop {} {\bibfield  {journal}
  {\bibinfo  {journal} {Chem. Phys. Lett.}\ }\textbf {\bibinfo {volume}
  {152}},\ \bibinfo {pages} {382} (\bibinfo {year} {1988})}\BibitemShut
  {NoStop}%
\bibitem [{CCp()}]{CCpy-GitHub}%
  \BibitemOpen
  \href@noop {} {}\bibinfo {note} {K. Gururangan and P. Piecuch, ``CCpy: A
  Coupled-Cluster Package Written in Python,'' see
  https://github.com/piecuch-group/ccpy}\BibitemShut {NoStop}%
\bibitem [{\citenamefont {Hirao}\ and\ \citenamefont
  {Nakatsuji}(1982)}]{hirao}%
  \BibitemOpen
  \bibfield  {author} {\bibinfo {author} {\bibfnamefont {K.}~\bibnamefont
  {Hirao}}\ and\ \bibinfo {author} {\bibfnamefont {H.}~\bibnamefont
  {Nakatsuji}},\ }\href@noop {} {\bibfield  {journal} {\bibinfo  {journal} {J.
  Comput. Phys.}\ }\textbf {\bibinfo {volume} {45}},\ \bibinfo {pages} {246}
  (\bibinfo {year} {1982})}\BibitemShut {NoStop}%
\bibitem [{\citenamefont {Davidson}(1975)}]{dav}%
  \BibitemOpen
  \bibfield  {author} {\bibinfo {author} {\bibfnamefont {E.~R.}\ \bibnamefont
  {Davidson}},\ }\href@noop {} {\bibfield  {journal} {\bibinfo  {journal} {J.
  Comput. Phys.}\ }\textbf {\bibinfo {volume} {17}},\ \bibinfo {pages} {87}
  (\bibinfo {year} {1975})}\BibitemShut {NoStop}%
\bibitem [{\citenamefont {Matthews}\ and\ \citenamefont
  {Stanton}(2016)}]{eomccsdta}%
  \BibitemOpen
  \bibfield  {author} {\bibinfo {author} {\bibfnamefont {D.~A.}\ \bibnamefont
  {Matthews}}\ and\ \bibinfo {author} {\bibfnamefont {J.~F.}\ \bibnamefont
  {Stanton}},\ }\href@noop {} {\bibfield  {journal} {\bibinfo  {journal} {J.
  Chem. Phys.}\ }\textbf {\bibinfo {volume} {145}},\ \bibinfo {pages} {124102}
  (\bibinfo {year} {2016})}\BibitemShut {NoStop}%
\bibitem [{\citenamefont {Dunning}(1989)}]{ccpvnz}%
  \BibitemOpen
  \bibfield  {author} {\bibinfo {author} {\bibfnamefont {T.~H.}\ \bibnamefont
  {Dunning}, \bibfnamefont {Jr.}},\ }\href@noop {} {\bibfield  {journal}
  {\bibinfo  {journal} {J. Chem. Phys.}\ }\textbf {\bibinfo {volume} {90}},\
  \bibinfo {pages} {1007} (\bibinfo {year} {1989})}\BibitemShut {NoStop}%
\bibitem [{\citenamefont {Kendall}, \citenamefont {Dunning},\ and\
  \citenamefont {Harrison}(1992)}]{augccpvnz}%
  \BibitemOpen
  \bibfield  {author} {\bibinfo {author} {\bibfnamefont {R.~A.}\ \bibnamefont
  {Kendall}}, \bibinfo {author} {\bibfnamefont {T.~H.}\ \bibnamefont {Dunning},
  \bibfnamefont {Jr.}}, \ and\ \bibinfo {author} {\bibfnamefont {R.~J.}\
  \bibnamefont {Harrison}},\ }\href@noop {} {\bibfield  {journal} {\bibinfo
  {journal} {J. Chem. Phys.}\ }\textbf {\bibinfo {volume} {96}},\ \bibinfo
  {pages} {6769} (\bibinfo {year} {1992})}\BibitemShut {NoStop}%
\bibitem [{\citenamefont {Marie}\ \emph {et~al.}(2024)\citenamefont {Marie},
  \citenamefont {Romaniello}, \citenamefont {Blase},\ and\ \citenamefont
  {Loos}}]{loos-ppbse}%
  \BibitemOpen
  \bibfield  {author} {\bibinfo {author} {\bibfnamefont {A.}~\bibnamefont
  {Marie}}, \bibinfo {author} {\bibfnamefont {P.}~\bibnamefont {Romaniello}},
  \bibinfo {author} {\bibfnamefont {X.}~\bibnamefont {Blase}}, \ and\ \bibinfo
  {author} {\bibfnamefont {P.-F.}\ \bibnamefont {Loos}},\ }\href@noop {}
  {\enquote {\bibinfo {title} {Anomalous propagators and the particle-particle
  channel: {B}ethe--{S}alpeter equation},}\ } (\bibinfo {year} {2024}),\
  \Eprint {http://arxiv.org/abs/2411.13167} {arXiv:2411.13167
  [physics.chem-ph]} \BibitemShut {NoStop}%
\bibitem [{\citenamefont {Huron}, \citenamefont {Malrieu},\ and\ \citenamefont
  {Rancurel}(1973)}]{sci_3}%
  \BibitemOpen
  \bibfield  {author} {\bibinfo {author} {\bibfnamefont {B.}~\bibnamefont
  {Huron}}, \bibinfo {author} {\bibfnamefont {J.~P.}\ \bibnamefont {Malrieu}},
  \ and\ \bibinfo {author} {\bibfnamefont {P.}~\bibnamefont {Rancurel}},\
  }\href@noop {} {\bibfield  {journal} {\bibinfo  {journal} {J. Chem. Phys.}\
  }\textbf {\bibinfo {volume} {58}},\ \bibinfo {pages} {5745} (\bibinfo {year}
  {1973})}\BibitemShut {NoStop}%
\bibitem [{\citenamefont {Garniron}\ \emph {et~al.}(2017)\citenamefont
  {Garniron}, \citenamefont {Scemama}, \citenamefont {Loos},\ and\
  \citenamefont {Caffarel}}]{cipsi_1}%
  \BibitemOpen
  \bibfield  {author} {\bibinfo {author} {\bibfnamefont {Y.}~\bibnamefont
  {Garniron}}, \bibinfo {author} {\bibfnamefont {A.}~\bibnamefont {Scemama}},
  \bibinfo {author} {\bibfnamefont {P.-F.}\ \bibnamefont {Loos}}, \ and\
  \bibinfo {author} {\bibfnamefont {M.}~\bibnamefont {Caffarel}},\ }\href@noop
  {} {\bibfield  {journal} {\bibinfo  {journal} {J. Chem. Phys.}\ }\textbf
  {\bibinfo {volume} {147}},\ \bibinfo {pages} {034101} (\bibinfo {year}
  {2017})}\BibitemShut {NoStop}%
\bibitem [{\citenamefont {Garniron}\ \emph {et~al.}(2019)\citenamefont
  {Garniron}, \citenamefont {Applencourt}, \citenamefont {Gasperich},
  \citenamefont {Benali}, \citenamefont {Ferte}, \citenamefont {Paquier},
  \citenamefont {Pradines}, \citenamefont {Assaraf}, \citenamefont {Reinhardt},
  \citenamefont {Toulouse}, \citenamefont {Barbaresco}, \citenamefont {Renon},
  \citenamefont {David}, \citenamefont {Malrieu}, \citenamefont {Véril},
  \citenamefont {Caffarel}, \citenamefont {Loos}, \citenamefont {Giner},\ and\
  \citenamefont {Scemama}}]{cipsi_2}%
  \BibitemOpen
  \bibfield  {author} {\bibinfo {author} {\bibfnamefont {Y.}~\bibnamefont
  {Garniron}}, \bibinfo {author} {\bibfnamefont {T.}~\bibnamefont
  {Applencourt}}, \bibinfo {author} {\bibfnamefont {K.}~\bibnamefont
  {Gasperich}}, \bibinfo {author} {\bibfnamefont {A.}~\bibnamefont {Benali}},
  \bibinfo {author} {\bibfnamefont {A.}~\bibnamefont {Ferte}}, \bibinfo
  {author} {\bibfnamefont {J.}~\bibnamefont {Paquier}}, \bibinfo {author}
  {\bibfnamefont {B.}~\bibnamefont {Pradines}}, \bibinfo {author}
  {\bibfnamefont {R.}~\bibnamefont {Assaraf}}, \bibinfo {author} {\bibfnamefont
  {P.}~\bibnamefont {Reinhardt}}, \bibinfo {author} {\bibfnamefont
  {J.}~\bibnamefont {Toulouse}}, \bibinfo {author} {\bibfnamefont
  {P.}~\bibnamefont {Barbaresco}}, \bibinfo {author} {\bibfnamefont
  {N.}~\bibnamefont {Renon}}, \bibinfo {author} {\bibfnamefont
  {G.}~\bibnamefont {David}}, \bibinfo {author} {\bibfnamefont {J.-P.}\
  \bibnamefont {Malrieu}}, \bibinfo {author} {\bibfnamefont {M.}~\bibnamefont
  {Véril}}, \bibinfo {author} {\bibfnamefont {M.}~\bibnamefont {Caffarel}},
  \bibinfo {author} {\bibfnamefont {P.-F.}\ \bibnamefont {Loos}}, \bibinfo
  {author} {\bibfnamefont {E.}~\bibnamefont {Giner}}, \ and\ \bibinfo {author}
  {\bibfnamefont {A.}~\bibnamefont {Scemama}},\ }\href@noop {} {\bibfield
  {journal} {\bibinfo  {journal} {J. Chem. Theory Comput.}\ }\textbf {\bibinfo
  {volume} {15}},\ \bibinfo {pages} {3591} (\bibinfo {year}
  {2019})}\BibitemShut {NoStop}%
\bibitem [{\citenamefont {McConkey}\ \emph {et~al.}(1994)\citenamefont
  {McConkey}, \citenamefont {Dawber}, \citenamefont {Avaldi}, \citenamefont
  {MacDonald}, \citenamefont {King},\ and\ \citenamefont {Hall}}]{dip_cl2}%
  \BibitemOpen
  \bibfield  {author} {\bibinfo {author} {\bibfnamefont {A.~G.}\ \bibnamefont
  {McConkey}}, \bibinfo {author} {\bibfnamefont {G.}~\bibnamefont {Dawber}},
  \bibinfo {author} {\bibfnamefont {L.}~\bibnamefont {Avaldi}}, \bibinfo
  {author} {\bibfnamefont {M.~A.}\ \bibnamefont {MacDonald}}, \bibinfo {author}
  {\bibfnamefont {G.~C.}\ \bibnamefont {King}}, \ and\ \bibinfo {author}
  {\bibfnamefont {R.~I.}\ \bibnamefont {Hall}},\ }\href@noop {} {\bibfield
  {journal} {\bibinfo  {journal} {J. Phys. B: At. Mol. Opt. Phys.}\ }\textbf
  {\bibinfo {volume} {27}},\ \bibinfo {pages} {271} (\bibinfo {year}
  {1994})}\BibitemShut {NoStop}%
\bibitem [{\citenamefont {Fleig}\ \emph {et~al.}(2008)\citenamefont {Fleig},
  \citenamefont {Edvardsson}, \citenamefont {Banks},\ and\ \citenamefont
  {Eland}}]{dip_br2}%
  \BibitemOpen
  \bibfield  {author} {\bibinfo {author} {\bibfnamefont {T.}~\bibnamefont
  {Fleig}}, \bibinfo {author} {\bibfnamefont {D.}~\bibnamefont {Edvardsson}},
  \bibinfo {author} {\bibfnamefont {S.~T.}\ \bibnamefont {Banks}}, \ and\
  \bibinfo {author} {\bibfnamefont {J.~H.}\ \bibnamefont {Eland}},\ }\href@noop
  {} {\bibfield  {journal} {\bibinfo  {journal} {Chem. Phys.}\ }\textbf
  {\bibinfo {volume} {343}},\ \bibinfo {pages} {270} (\bibinfo {year}
  {2008})}\BibitemShut {NoStop}%
\bibitem [{\citenamefont {Eland}(2003)}]{dip_hbr}%
  \BibitemOpen
  \bibfield  {author} {\bibinfo {author} {\bibfnamefont {J.~H.}\ \bibnamefont
  {Eland}},\ }\href@noop {} {\bibfield  {journal} {\bibinfo  {journal} {Chem.
  Phys.}\ }\textbf {\bibinfo {volume} {294}},\ \bibinfo {pages} {171} (\bibinfo
  {year} {2003})}\BibitemShut {NoStop}%
\bibitem [{\citenamefont {Woon}\ and\ \citenamefont {Dunning}(1993)}]{ccpvnz3}%
  \BibitemOpen
  \bibfield  {author} {\bibinfo {author} {\bibfnamefont {D.~E.}\ \bibnamefont
  {Woon}}\ and\ \bibinfo {author} {\bibfnamefont {T.~H.}\ \bibnamefont
  {Dunning}, \bibfnamefont {Jr.}},\ }\href@noop {} {\bibfield  {journal}
  {\bibinfo  {journal} {J. Chem. Phys.}\ }\textbf {\bibinfo {volume} {98}},\
  \bibinfo {pages} {1358} (\bibinfo {year} {1993})}\BibitemShut {NoStop}%
\bibitem [{\citenamefont {Wilson}\ \emph {et~al.}(1999)\citenamefont {Wilson},
  \citenamefont {Woon}, \citenamefont {Peterson},\ and\ \citenamefont
  {Dunning}}]{ccpvnz9}%
  \BibitemOpen
  \bibfield  {author} {\bibinfo {author} {\bibfnamefont {A.~K.}\ \bibnamefont
  {Wilson}}, \bibinfo {author} {\bibfnamefont {D.~E.}\ \bibnamefont {Woon}},
  \bibinfo {author} {\bibfnamefont {K.~A.}\ \bibnamefont {Peterson}}, \ and\
  \bibinfo {author} {\bibfnamefont {T.~H.}\ \bibnamefont {Dunning},
  \bibfnamefont {Jr.}},\ }\href@noop {} {\bibfield  {journal} {\bibinfo
  {journal} {J. Chem. Phys.}\ }\textbf {\bibinfo {volume} {110}},\ \bibinfo
  {pages} {7667} (\bibinfo {year} {1999})}\BibitemShut {NoStop}%
\bibitem [{\citenamefont {Marie}\ and\ \citenamefont {Loos}(2024)}]{loos-ip}%
  \BibitemOpen
  \bibfield  {author} {\bibinfo {author} {\bibfnamefont {A.}~\bibnamefont
  {Marie}}\ and\ \bibinfo {author} {\bibfnamefont {P.-F.}\ \bibnamefont
  {Loos}},\ }\href@noop {} {\bibfield  {journal} {\bibinfo  {journal} {J. Chem.
  Theory Comput.}\ }\textbf {\bibinfo {volume} {20}},\ \bibinfo {pages} {4751}
  (\bibinfo {year} {2024})}\BibitemShut {NoStop}%
\bibitem [{\citenamefont {Huber}\ and\ \citenamefont
  {Herzberg}(1979)}]{herzberg4}%
  \BibitemOpen
  \bibfield  {author} {\bibinfo {author} {\bibfnamefont {K.~P.}\ \bibnamefont
  {Huber}}\ and\ \bibinfo {author} {\bibfnamefont {G.}~\bibnamefont
  {Herzberg}},\ }\href@noop {} {\emph {\bibinfo {title} {Molecular Spectra and
  Molecular Structure: Constants of Diatomic Molecules}}}\ (\bibinfo
  {publisher} {Van Nostrand Reinhold},\ \bibinfo {address} {New York},\
  \bibinfo {year} {1979})\BibitemShut {NoStop}%
\bibitem [{\citenamefont {Sun}\ \emph {et~al.}(2018)\citenamefont {Sun},
  \citenamefont {Berkelbach}, \citenamefont {Blunt}, \citenamefont {Booth},
  \citenamefont {Guo}, \citenamefont {Li}, \citenamefont {Liu}, \citenamefont
  {McClain}, \citenamefont {Sayfutyarova}, \citenamefont {Sharma},
  \citenamefont {Wouters},\ and\ \citenamefont {Chan}}]{pyscf1}%
  \BibitemOpen
  \bibfield  {author} {\bibinfo {author} {\bibfnamefont {Q.}~\bibnamefont
  {Sun}}, \bibinfo {author} {\bibfnamefont {T.~C.}\ \bibnamefont {Berkelbach}},
  \bibinfo {author} {\bibfnamefont {N.~S.}\ \bibnamefont {Blunt}}, \bibinfo
  {author} {\bibfnamefont {G.~H.}\ \bibnamefont {Booth}}, \bibinfo {author}
  {\bibfnamefont {S.}~\bibnamefont {Guo}}, \bibinfo {author} {\bibfnamefont
  {Z.}~\bibnamefont {Li}}, \bibinfo {author} {\bibfnamefont {J.}~\bibnamefont
  {Liu}}, \bibinfo {author} {\bibfnamefont {J.~D.}\ \bibnamefont {McClain}},
  \bibinfo {author} {\bibfnamefont {E.~R.}\ \bibnamefont {Sayfutyarova}},
  \bibinfo {author} {\bibfnamefont {S.}~\bibnamefont {Sharma}}, \bibinfo
  {author} {\bibfnamefont {S.}~\bibnamefont {Wouters}}, \ and\ \bibinfo
  {author} {\bibfnamefont {G.~K.-L.}\ \bibnamefont {Chan}},\ }\href@noop {}
  {\bibfield  {journal} {\bibinfo  {journal} {WIREs Comput. Mol. Sci.}\
  }\textbf {\bibinfo {volume} {8}},\ \bibinfo {pages} {e1340} (\bibinfo {year}
  {2018})}\BibitemShut {NoStop}%
\bibitem [{\citenamefont {Sun}\ \emph {et~al.}(2020)\citenamefont {Sun},
  \citenamefont {Zhang}, \citenamefont {Banerjee}, \citenamefont {Bao},
  \citenamefont {Barbry}, \citenamefont {Blunt}, \citenamefont {Bogdanov},
  \citenamefont {Booth}, \citenamefont {Chen}, \citenamefont {Cui},
  \citenamefont {Eriksen}, \citenamefont {Gao}, \citenamefont {Guo},
  \citenamefont {Hermann}, \citenamefont {Hermes}, \citenamefont {Koh},
  \citenamefont {Koval}, \citenamefont {Lehtola}, \citenamefont {Li},
  \citenamefont {Liu}, \citenamefont {Mardirossian}, \citenamefont {McClain},
  \citenamefont {Motta}, \citenamefont {Mussard}, \citenamefont {Pham},
  \citenamefont {Pulkin}, \citenamefont {Purwanto}, \citenamefont {Robinson},
  \citenamefont {Ronca}, \citenamefont {Sayfutyarova}, \citenamefont
  {Scheurer}, \citenamefont {Schurkus}, \citenamefont {Smith}, \citenamefont
  {Sun}, \citenamefont {Sun}, \citenamefont {Upadhyay}, \citenamefont {Wagner},
  \citenamefont {Wang}, \citenamefont {White}, \citenamefont {Whitfield},
  \citenamefont {Williamson}, \citenamefont {Wouters}, \citenamefont {Yang},
  \citenamefont {Yu}, \citenamefont {Zhu}, \citenamefont {Berkelbach},
  \citenamefont {Sharma}, \citenamefont {Sokolov},\ and\ \citenamefont
  {Chan}}]{pyscf2}%
  \BibitemOpen
  \bibfield  {author} {\bibinfo {author} {\bibfnamefont {Q.}~\bibnamefont
  {Sun}}, \bibinfo {author} {\bibfnamefont {X.}~\bibnamefont {Zhang}}, \bibinfo
  {author} {\bibfnamefont {S.}~\bibnamefont {Banerjee}}, \bibinfo {author}
  {\bibfnamefont {P.}~\bibnamefont {Bao}}, \bibinfo {author} {\bibfnamefont
  {M.}~\bibnamefont {Barbry}}, \bibinfo {author} {\bibfnamefont {N.~S.}\
  \bibnamefont {Blunt}}, \bibinfo {author} {\bibfnamefont {N.~A.}\ \bibnamefont
  {Bogdanov}}, \bibinfo {author} {\bibfnamefont {G.~H.}\ \bibnamefont {Booth}},
  \bibinfo {author} {\bibfnamefont {J.}~\bibnamefont {Chen}}, \bibinfo {author}
  {\bibfnamefont {Z.-H.}\ \bibnamefont {Cui}}, \bibinfo {author} {\bibfnamefont
  {J.~J.}\ \bibnamefont {Eriksen}}, \bibinfo {author} {\bibfnamefont
  {Y.}~\bibnamefont {Gao}}, \bibinfo {author} {\bibfnamefont {S.}~\bibnamefont
  {Guo}}, \bibinfo {author} {\bibfnamefont {J.}~\bibnamefont {Hermann}},
  \bibinfo {author} {\bibfnamefont {M.~R.}\ \bibnamefont {Hermes}}, \bibinfo
  {author} {\bibfnamefont {K.}~\bibnamefont {Koh}}, \bibinfo {author}
  {\bibfnamefont {P.}~\bibnamefont {Koval}}, \bibinfo {author} {\bibfnamefont
  {S.}~\bibnamefont {Lehtola}}, \bibinfo {author} {\bibfnamefont
  {Z.}~\bibnamefont {Li}}, \bibinfo {author} {\bibfnamefont {J.}~\bibnamefont
  {Liu}}, \bibinfo {author} {\bibfnamefont {N.}~\bibnamefont {Mardirossian}},
  \bibinfo {author} {\bibfnamefont {J.~D.}\ \bibnamefont {McClain}}, \bibinfo
  {author} {\bibfnamefont {M.}~\bibnamefont {Motta}}, \bibinfo {author}
  {\bibfnamefont {B.}~\bibnamefont {Mussard}}, \bibinfo {author} {\bibfnamefont
  {H.~Q.}\ \bibnamefont {Pham}}, \bibinfo {author} {\bibfnamefont
  {A.}~\bibnamefont {Pulkin}}, \bibinfo {author} {\bibfnamefont
  {W.}~\bibnamefont {Purwanto}}, \bibinfo {author} {\bibfnamefont {P.~J.}\
  \bibnamefont {Robinson}}, \bibinfo {author} {\bibfnamefont {E.}~\bibnamefont
  {Ronca}}, \bibinfo {author} {\bibfnamefont {E.~R.}\ \bibnamefont
  {Sayfutyarova}}, \bibinfo {author} {\bibfnamefont {M.}~\bibnamefont
  {Scheurer}}, \bibinfo {author} {\bibfnamefont {H.~F.}\ \bibnamefont
  {Schurkus}}, \bibinfo {author} {\bibfnamefont {J.~E.~T.}\ \bibnamefont
  {Smith}}, \bibinfo {author} {\bibfnamefont {C.}~\bibnamefont {Sun}}, \bibinfo
  {author} {\bibfnamefont {S.-N.}\ \bibnamefont {Sun}}, \bibinfo {author}
  {\bibfnamefont {S.}~\bibnamefont {Upadhyay}}, \bibinfo {author}
  {\bibfnamefont {L.~K.}\ \bibnamefont {Wagner}}, \bibinfo {author}
  {\bibfnamefont {X.}~\bibnamefont {Wang}}, \bibinfo {author} {\bibfnamefont
  {A.}~\bibnamefont {White}}, \bibinfo {author} {\bibfnamefont {J.~D.}\
  \bibnamefont {Whitfield}}, \bibinfo {author} {\bibfnamefont {M.~J.}\
  \bibnamefont {Williamson}}, \bibinfo {author} {\bibfnamefont
  {S.}~\bibnamefont {Wouters}}, \bibinfo {author} {\bibfnamefont
  {J.}~\bibnamefont {Yang}}, \bibinfo {author} {\bibfnamefont {J.~M.}\
  \bibnamefont {Yu}}, \bibinfo {author} {\bibfnamefont {T.}~\bibnamefont
  {Zhu}}, \bibinfo {author} {\bibfnamefont {T.~C.}\ \bibnamefont {Berkelbach}},
  \bibinfo {author} {\bibfnamefont {S.}~\bibnamefont {Sharma}}, \bibinfo
  {author} {\bibfnamefont {A.~Y.}\ \bibnamefont {Sokolov}}, \ and\ \bibinfo
  {author} {\bibfnamefont {G.~K.-L.}\ \bibnamefont {Chan}},\ }\href@noop {}
  {\bibfield  {journal} {\bibinfo  {journal} {J. Chem. Phys.}\ }\textbf
  {\bibinfo {volume} {153}},\ \bibinfo {pages} {024109} (\bibinfo {year}
  {2020})}\BibitemShut {NoStop}%
\bibitem [{\citenamefont {Cheng}\ and\ \citenamefont {Gauss}(2011)}]{sfx2c1e}%
  \BibitemOpen
  \bibfield  {author} {\bibinfo {author} {\bibfnamefont {L.}~\bibnamefont
  {Cheng}}\ and\ \bibinfo {author} {\bibfnamefont {J.}~\bibnamefont {Gauss}},\
  }\href@noop {} {\bibfield  {journal} {\bibinfo  {journal} {J. Chem. Phys.}\
  }\textbf {\bibinfo {volume} {135}},\ \bibinfo {pages} {084114} (\bibinfo
  {year} {2011})}\BibitemShut {NoStop}%
\bibitem [{\citenamefont {Oliphant}\ and\ \citenamefont
  {Adamowicz}(1992)}]{semi0b}%
  \BibitemOpen
  \bibfield  {author} {\bibinfo {author} {\bibfnamefont {N.}~\bibnamefont
  {Oliphant}}\ and\ \bibinfo {author} {\bibfnamefont {L.}~\bibnamefont
  {Adamowicz}},\ }\href@noop {} {\bibfield  {journal} {\bibinfo  {journal} {J.
  Chem. Phys.}\ }\textbf {\bibinfo {volume} {96}},\ \bibinfo {pages} {3739}
  (\bibinfo {year} {1992})}\BibitemShut {NoStop}%
\bibitem [{\citenamefont {Piecuch}, \citenamefont {Oliphant},\ and\
  \citenamefont {Adamowicz}(1993)}]{semi2}%
  \BibitemOpen
  \bibfield  {author} {\bibinfo {author} {\bibfnamefont {P.}~\bibnamefont
  {Piecuch}}, \bibinfo {author} {\bibfnamefont {N.}~\bibnamefont {Oliphant}}, \
  and\ \bibinfo {author} {\bibfnamefont {L.}~\bibnamefont {Adamowicz}},\
  }\href@noop {} {\bibfield  {journal} {\bibinfo  {journal} {J. Chem. Phys.}\
  }\textbf {\bibinfo {volume} {99}},\ \bibinfo {pages} {1875} (\bibinfo {year}
  {1993})}\BibitemShut {NoStop}%
\bibitem [{\citenamefont {Piecuch}, \citenamefont {Kucharski},\ and\
  \citenamefont {Bartlett}(1999)}]{semi4}%
  \BibitemOpen
  \bibfield  {author} {\bibinfo {author} {\bibfnamefont {P.}~\bibnamefont
  {Piecuch}}, \bibinfo {author} {\bibfnamefont {S.~A.}\ \bibnamefont
  {Kucharski}}, \ and\ \bibinfo {author} {\bibfnamefont {R.~J.}\ \bibnamefont
  {Bartlett}},\ }\href@noop {} {\bibfield  {journal} {\bibinfo  {journal} {J.
  Chem. Phys.}\ }\textbf {\bibinfo {volume} {110}},\ \bibinfo {pages} {6103}
  (\bibinfo {year} {1999})}\BibitemShut {NoStop}%
\bibitem [{\citenamefont {Surjuse}\ \emph {et~al.}(2022)\citenamefont
  {Surjuse}, \citenamefont {Chamoli}, \citenamefont {Nayak},\ and\
  \citenamefont {Dutta}}]{4c-eomcc-fns}%
  \BibitemOpen
  \bibfield  {author} {\bibinfo {author} {\bibfnamefont {K.}~\bibnamefont
  {Surjuse}}, \bibinfo {author} {\bibfnamefont {S.}~\bibnamefont {Chamoli}},
  \bibinfo {author} {\bibfnamefont {M.~K.}\ \bibnamefont {Nayak}}, \ and\
  \bibinfo {author} {\bibfnamefont {A.~K.}\ \bibnamefont {Dutta}},\ }\href@noop
  {} {\bibfield  {journal} {\bibinfo  {journal} {J. Chem. Phys.}\ }\textbf
  {\bibinfo {volume} {157}},\ \bibinfo {pages} {204106} (\bibinfo {year}
  {2022})}\BibitemShut {NoStop}%
\end{thebibliography}%

\newpage
\clearpage
\onecolumngrid

\begin{table*}[h]
\caption{
\label{table1} 
The one- and two-body components of the similarity-transformed Hamiltonian 
expressed in terms of the one-, two-, and three-body cluster amplitudes,
$t_{a}^{i}$, $t_{ab}^{ij}$, and $t_{abc}^{ijk}$, respectively,
and matrix elements of the Fock and two-electron
interaction operators, denoted as $f_{p}^{q} \equiv \langle p | f | q \rangle$
and $v_{pq}^{rs} \equiv \langle pq | v | rs \rangle - \langle pq | v | sr \rangle$.
In the case of the DIP-EOMCCSDT(4$h$-2$p$) method, the $t_{a}^{i}$, $t_{ab}^{ij}$, and $t_{abc}^{ijk}$
values are obtained with CCSDT.
In the case of the DIP-EOMCCSD(T)(a)(4$h$-2$p$) approach, they are
replaced by their $\tilde{t}_{a}^{i}$, $\tilde{t}_{ab}^{ij}$, and $\tilde{t}_{abc}^{ijk}$ 
counterparts corresponding to the $\tilde{T}_{1}$, $\tilde{T}_{2}$, and $\tilde{T}_{3}$ 
operators defined by Eqs.\ (\ref{t1pert})--(\ref{t3pert}).
}
\begin{ruledtabular}
\begin{tabular}{cc}
Component of $\overline{H}_{N}$ & Expression\footnotemark[1] \\
\hline
%
%
%
\vspace{0.2em}
$\bar{h}_{m}^{e}$ & $f_{m}^{e} + v_{im}^{ae} t_{e}^{m}$ \vspace{0.3em}\\
$\bar{h}_{j}^{i}$ & $f_{j}^{i} + \bar{h}_{j}^{e} t_{e}^{i} + v_{jm}^{ie} t_{e}^{m} + \tfrac{1}{2} v_{jn}^{ef} t_{ef}^{in}$ \vspace{0.3em}\\
$\bar{h}_{a}^{b}$ & $f_{a}^{b} - \bar{h}_{m}^{b} t_{a}^{m} + v_{am}^{be} t_{e}^{m} - \tfrac{1}{2} v_{mn}^{bf} t_{af}^{mn}$ \vspace{0.3em}\\
$\bar{h}_{mn}^{ef}$ & $v_{mn}^{ef}$ \vspace{0.3em}\\
$\bar{h}_{am}^{ef}$ & $v_{am}^{ef} - v_{mn}^{fe} t_{a}^{n}$ \vspace{0.3em}\\
$\bar{h}_{mn}^{ie}$ & $v_{mn}^{ie} + v_{mn}^{fe} t_{f}^{i}$ \vspace{0.3em}\\
$\bar{h}_{ab}^{ef}$ & $v_{ab}^{ef} + \tfrac{1}{2} v_{mn}^{ef} \tau_{ab}^{mn} - \mathscr{A}_{ab} v_{am}^{ef} t_{b}^{m}$ \vspace{0.3em}\\
$\bar{h}_{mn}^{ij}$ & $v_{mn}^{ij} + \tfrac{1}{2} v_{mn}^{ef} \tau_{ef}^{ij} + \mathscr{A}^{ij} v_{nm}^{je} t_{e}^{i}$ \vspace{0.3em}\\
$\bar{h}_{am}^{ie}$ & $v_{am}^{ie} + v_{am}^{fe} t_{f}^{i} - \bar{h}_{nm}^{ie} t_{a}^{n} + v_{mn}^{ef} t_{af}^{in}$ \vspace{0.3em}\\
$\bar{h}_{am}^{ij}$ & $v_{am}^{ij} + \bar{h}_{m}^{e} t_{ae}^{ij} - \bar{h}_{nm}^{ij} t_{a}^{n} + \tfrac{1}{2} v_{am}^{ef} t_{ef}^{ij} + \mathscr{A}^{ij}(\bar{h}_{mn}^{jf} t_{af}^{in} + {\chi^{\prime}}_{am}^{ie} t_{e}^{j}) + v_{mn}^{ef} t_{aef}^{ijn}$ \vspace{0.3em}\\
$\bar{h}_{ab}^{ie}$ & $v_{ab}^{ie} - \bar{h}_{m}^{e} t_{ab}^{im} + v_{ab}^{ef} t_{f}^{i} + \tfrac{1}{2} \bar{h}_{mn}^{ie} t_{ab}^{mn} - \mathscr{A}_{ab} (\chi_{am}^{ie} t_{b}^{m} - v_{bn}^{ef} t_{af}^{in}) - v_{mn}^{ef} t_{abf}^{imn}$ \\
%
%
${\chi^{\prime}}_{am}^{ie}$ & $v_{am}^{ie} + \tfrac{1}{2} v_{am}^{ef} t_{e}^{i}$ \vspace{0.3em}\\
$\chi_{am}^{ie}$ & $\bar{h}_{am}^{ie} + \tfrac{1}{2} \bar{h}_{nm}^{ie} t_{a}^{n}$ \vspace{0.3em}\\
$\tau_{ab}^{ij}$ & $t_{ab}^{ij} + \mathscr{A}^{ij} t_{a}^{i} t_{b}^{j}$ \\
\end{tabular}
\end{ruledtabular}
\footnotetext[1]{
In each expression, summation is carried out over repeated upper and lower indices.
}
\end{table*}

\begin{table*}[h]
\caption{
\label{table2} 
The intermediates entering Eqs.\ (\ref{eq3h1p}) and (\ref{eq4h2p})
that are introduced in order to evaluate the contributions to the
DIP-EOMCCSDT(4$h$-2$p$) and DIP-EOMCCSD(T)(a)(4$h$-2$p$)
equations due to the three- and four-body components 
of the similarity-transformed Hamiltonian.
}
\begin{ruledtabular}
\begin{tabular}{cc}
Intermediate & Expression\footnotemark[1] \\
\hline
%
%
%
\vspace{0.2em}
%
%
${I^{\prime}}^{ie}(\mu)$ & $\tfrac{1}{2}\bar{h}_{mn}^{ie} r^{mn}(\mu) - \tfrac{1}{2}\bar{h}_{mn}^{fe} r_{\phantom{ab}f}^{inm}(\mu)$ \vspace{0.3em}\\
$I^{ie}(\mu)$\footnotemark[2] & $\tfrac{1}{2}\bar{h}_{mn}^{ie} r^{mn}(\mu) - \tfrac{1}{2}\bar{h}_{mn}^{fe} r_{\phantom{ab}f}^{inm}(\mu) - \bar{h}_{m}^{e} r^{im}(\mu)$ \vspace{0.3em}\\
$I^{ef}(\mu)$ & $\tfrac{1}{2} \bar{h}_{mn}^{ef} r^{mn}(\mu)$ \vspace{0.3em}\\
$I_{\phantom{ab}m}^{ijk}(\mu)$ & $\mathscr{A}^{ijk}[\tfrac{1}{2} \bar{h}_{nm}^{ke} r_{\phantom{ab}e}^{ijm}(\mu) - \tfrac{1}{2} \bar{h}_{nm}^{ik} r^{nj}(\mu) + \tfrac{1}{12} \bar{h}_{mn}^{ef} r_{\phantom{ab}ef}^{ijkn}(\mu)]$ \vspace{0.3em}\\
$I_{\phantom{ab}c}^{ije}(\mu)$ & $\bar{h}_{cm}^{fe} r_{\phantom{ab}e}^{ijm}(\mu) + \tfrac{1}{2} I^{fe}(\mu) t_{cf}^{ij} + \mathscr{A}^{ij}[\bar{h}_{cm}^{ie} r^{mj}(\mu) + \tfrac{1}{2} \bar{h}_{nm}^{ie} r_{\phantom{ab}c}^{njm}(\mu)] - \tfrac{1}{2} \bar{h}_{mn}^{ef} r_{\phantom{ab}cf}^{ijmn}(\mu)$ \vspace{0.3em}\\
$I_{\phantom{ab}e}^{ijm}(\mu)$\footnotemark[3] & $\bar{h}_{mn}^{ef} r_{\phantom{ab}f}^{ijn}(\mu) - \mathscr{A}^{ij} \bar{h}_{nm}^{je} r^{in}(\mu)$ \vspace{0.3em}\\
$I_{\phantom{ab}c}^{efk}(\mu)$\footnotemark[4] & $\tfrac{1}{2} \bar{h}_{mn}^{ef} r_{\phantom{ab}c}^{mnk}(\mu) + \bar{h}_{cm}^{ef} r^{km}(\mu)$ \\
\end{tabular}
\end{ruledtabular}
\footnotetext[1]{
In each expression, summation is carried out over repeated upper and lower indices.
}
\footnotetext[2]{
In the expression for $I^{ie}(\mu)$ used in
DIP-EOMCCSD(T)(a)(4$h$-2$p$), $\bar{h}_{mn}^{ie}$ and $\bar{h}_{m}^{e}$
are replaced by $v_{mn}^{ie}$ and $f_{m}^{e}$, respectively.
}
\footnotetext[3]{
In the expression for $I_{\phantom{ab}e}^{ijm}(\mu)$ used in
DIP-EOMCCSD(T)(a)(4$h$-2$p$), $\bar{h}_{mn}^{je}$
is replaced by $v_{mn}^{je}$.
}
\footnotetext[4]{
In the expression for $I_{\phantom{ab}c}^{efk}(\mu)$ used in
DIP-EOMCCSD(T)(a)(4$h$-2$p$), $\bar{h}_{cm}^{ef}$
is replaced by $v_{cm}^{ef}$.
}
\end{table*}

\newpage
\clearpage

\squeezetable
\begin{table*}[h]
\caption{
\label{table3}
The vertical DIP energies, in eV, of \water{}, \methane{}, and BN
corresponding to the lowest singlet and triplet states of the dicationic
species obtained in the DIP-EOMCCSD(3$h$-1$p$), DIP-EOMCCSD(4$h$-2$p$), 
DIP-EOMCCSD(T)(a)(4$h$-2$p$), and DIP-EOMCCSDT(4$h$-2$p$) calculations, 
alongside the high-level benchmark DIPs taken from Ref.\ \onlinecite{loos-ppbse}
resulting from the CIPSI computations extrapolated to the exact, full CI, limit.
The DIP-EOMCC calculations used the RHF orbitals of the respective neutral systems.
All calculations employed the aug-cc-pVTZ basis set and the frozen-core
approximation was assumed in all post-RHF steps. 
}
\begin{ruledtabular}
\begin{tabular}{c c c c c c c c c c c}
Molecule & 
Dication State & 
\multicolumn{1}{c}{CCSD($3h\mbox{-}1p$)\footnotemark[1]} & 
\multicolumn{1}{c}{CCSD($4h\mbox{-}2p$)\footnotemark[2]} & 
\multicolumn{1}{c}{CCSD(T)(a)($4h\mbox{-}2p$)\footnotemark[3]} & 
\multicolumn{1}{c}{CCSDT($4h\mbox{-}2p$)\footnotemark[4]} & 
CIPSI\footnotemark[5] \\
\hline
%
\water{}\footnotemark[6] & \watertriplet{} & 41.06 & 40.04 & 40.26 & 40.27 & 40.29 \\ 
                         & \watersinglet{} & 42.04 & 41.18 & 41.39 & 41.40 & 41.43 \\ [1mm]
\methane{}\footnotemark[7] & \methanetriplet{} & 38.59 & 38.10 & 38.26 & 38.27 & 38.27 \\ 
                           & \methanesinglet{} & 39.29 & 38.80 & 38.96 & 38.97 & 38.98 \\ [1mm]
BN\footnotemark[8] & \bntriplet{} & 34.17 & 33.12 & 33.76 & 33.74 & 33.73 \\ 
                   & \bnsinglet{} & 35.33 & 34.31 & 34.98 & 34.98 & 34.98
\end{tabular}
\end{ruledtabular}
\footnotetext[1]{
The DIP-EOMCCSD(3$h$-1$p$) approach.
}
\footnotetext[2]{
The DIP-EOMCCSD(4$h$-2$p$) approach.
}
\footnotetext[3]{
The DIP-EOMCCSD(T)(a)(4$h$-2$p$) approach.
}
\footnotetext[4]{
The DIP-EOMCCSDT(4$h$-2$p$) approach.
}
\footnotetext[5]{
The results of CIPSI calculations extrapolated to the exact, full CI, limit, taken from Ref.\ \onlinecite{loos-ppbse}.
}
\footnotetext[6]{
The O--H bond length and the H--O--H bond angle in the $C_{2v}$-symmetric ground-state \water{}, optimized using
CC3/aug-cc-pVTZ and taken from Ref.\ \onlinecite{loos-ip}, are 0.9591 \AA{} and 103.2 degree, respectively.
}
\footnotetext[7]{
The C--H bond length in the $T_{d}$-symmetric ground-state \methane{}, optimized using 
CC3/aug-cc-pVTZ and taken from Ref.\ \onlinecite{loos-ip}, is 1.0879 \AA{}.
}
\footnotetext[8]{
The equilibrium B--N bond length in the ground-state BN, optimized using CC3/aug-cc-pVTZ and taken from Ref.\ \onlinecite{loos-ip}, is 1.2765 \AA{}.
}
\end{table*}

\squeezetable
\begin{table*}[h]
\caption{
\label{table4}
The vertical DIP energies, in eV, of Cl$_{\rm 2}$, Br$_{\rm 2}$, and HBr
corresponding to the experimental data
reported in Refs.\ \onlinecite{dip_cl2,dip_br2,dip_hbr} obtained in the DIP-EOMCCSD(3$h$-1$p$),
DIP-EOMCCSD(4$h$-2$p$), DIP-EOMCCSD(T)(a)(4$h$-2$p$), and DIP-EOMCCSDT(4$h$-2$p$) 
calculations using the cc-pVTZ and cc-pVQZ basis sets. Scalar-relativistic effects were
incorporated using the SFX2C-1e methodology of Ref.\ \onlinecite{sfx2c1e}.
All DIP-EOMCC calculations used the 
RHF orbitals of the respective neutral diatomics and the frozen-core approximation 
was assumed in all post-RHF steps. 
}
\begin{ruledtabular}
\begin{tabular}{c c c c c c c c c c c}
Molecule & 
Dication State & 
\multicolumn{2}{c}{CCSD($3h\mbox{-}1p$)\footnotemark[1]} & 
\multicolumn{2}{c}{CCSD($4h\mbox{-}2p$)\footnotemark[2]} & 
\multicolumn{2}{c}{CCSD(T)(a)($4h\mbox{-}2p$)\footnotemark[3]} & 
\multicolumn{2}{c}{CCSDT($4h\mbox{-}2p$)\footnotemark[4]} & 
Experiment\footnotemark[5] \\
\cline{3-4} 
\cline{5-6}
\cline{7-8}
\cline{9-10}
& & TZ & QZ & TZ & QZ & TZ & QZ & TZ & QZ & \\
\hline
%
$\rm{Cl_{2}}$\footnotemark[6] & $X\:{^{3}}\Sigma_{g}^{-}$ & 31.28 & 31.57 & 30.58 & 30.82 & 30.82 & 31.12 & 30.84 & 31.13 & 31.13 \\ 
			               & $a\:{^{1}}\Delta_{g}$     & 31.78 & 32.06 & 31.12 & 31.35 & 31.36 & 31.64 & 31.37 & 31.66 & 31.74 \\
			               & $b\:{^{1}}\Sigma_{g}^{+}$ & 32.16 & 32.45 & 31.51 & 31.74 & 31.74 & 32.03 & 31.76 & 32.05 & 32.12 \\
				       & $c\:{^{1}}\Sigma_{u}^{-}$ & 33.22 & 33.52 & 32.56 & 32.82 & 32.79 & 33.09 & 32.80 & 33.11 & 32.97 \\ [1mm]
$\rm{Br_{2}}$\footnotemark[7] & $X\:{^{3}}\Sigma_{g}^{-}$ & 28.47 & 28.68 & 27.92 & 28.12 & 28.12 & 28.35 & 28.13 & 28.37 & 28.53 \\
            			   & $a\:{^{1}}\Delta_{g}$     & 28.89 & 29.10 & 28.38 & 28.56 & 28.57 & 28.79 & 28.58 & 28.81 & 28.91 \\
            			   & $b\:{^{1}}\Sigma_{g}^{+}$ & 29.21 & 29.42 & 28.71 & 28.90 & 28.90 & 29.13 & 28.91 & 29.14 & 29.38 \\
				   & $c\:{^{1}}\Sigma_{u}^{-}$ & 29.93 & 30.16 & 29.43 & 29.63 & 29.61 & 29.85 & 29.62 & 29.87 & 30.3 \\ [1mm]
$\rm{HBr}$\footnotemark[8]    & $X\:{^{3}}\Sigma^{-}$ & 32.70 & 32.93 & 32.29 & 32.51 & 32.42 & 32.67 & 32.43 & 32.69 & 32.62 \\
        			       & $a\:{^{1}}\Delta $    & 34.06 & 34.25 & 33.69 & 33.86 & 33.82 & 34.02 & 33.83 & 34.04 & 33.95 \\
        			       & $b\:{^{1}}\Sigma^{+}$ & 35.31 & 35.50 & 34.95 & 35.12 & 35.07 & 35.27 & 35.09 & 35.28 & 35.19
\end{tabular}
\end{ruledtabular}
\footnotetext[1]{
The DIP-EOMCCSD(3$h$-1$p$) approach.
}
\footnotetext[2]{
The DIP-EOMCCSD(4$h$-2$p$) approach.
}
\footnotetext[3]{
The DIP-EOMCCSD(T)(a)(4$h$-2$p$) approach.
}
\footnotetext[4]{
The DIP-EOMCCSDT(4$h$-2$p$) approach.
}
\footnotetext[5]{
The experimentally determined DIP values taken
from Ref.\ \onlinecite{dip_cl2} for \Cltwo{}, Ref.\ \onlinecite{dip_br2} for \Brtwo{},
and Ref.\ \onlinecite{dip_hbr} for HBr.
}
\footnotetext[6]{
The equilibrium Cl--Cl bond length in the ground-state \Cltwo{}, taken from Ref.\ \onlinecite{herzberg4}, is 1.987 \AA.
}
\footnotetext[7]{
The equilibrium Br--Br bond length in the ground-state \Brtwo{}, taken from Ref.\ \onlinecite{herzberg4}, is 2.281 \AA.
}
\footnotetext[8]{
The equilibrium H--Br bond length in the ground-state HBr, taken from Ref.\ \onlinecite{herzberg4}, is 1.414 \AA.
}
\end{table*}
\end{document}